\pdfoutput=1

\PassOptionsToPackage{svgnames,x11names}{xcolor}
\documentclass[sigconf,nonacm]{acmart}
\usepackage[T1]{fontenc}
\usepackage[utf8]{inputenc}

\usepackage{graphicx}
\usepackage{makecell}

\usepackage{svg}
\usepackage{circledsteps}

\DeclareFontFamily{U}{FdSymbolA}{}
\DeclareFontShape{U}{FdSymbolA}{m}{n}{<-> s * FdSymbolA-Book}{}
\DeclareSymbolFont{fdbigsign}{U}{FdSymbolA}{m}{n}
\DeclareMathSymbol{\medblacksquare}{\mathrel}{fdbigsign}{"76}
\DeclareMathSymbol{\medblackdiamond}{\mathrel}{fdbigsign}{"84}
\DeclareMathSymbol{\medblackcircle}{\mathrel}{fdbigsign}{"63}
\DeclareMathSymbol{\medblackplus}{\mathrel}{fdbigsign}{"11}
\DeclareMathSymbol{\medblacktimes}{\mathrel}{fdbigsign}{"12}

\usepackage{times}
\usepackage[scaled=0.92]{helvet} 
\usepackage{tabularx}
\usepackage{colortbl}
\usepackage{array}
\usepackage{booktabs}
\usepackage{microtype}
   \microtypecontext{spacing=nonfrench}
\usepackage{paralist}
\usepackage{soul}
\usepackage{xspace}
\usepackage{tikz}
   \usetikzlibrary{positioning}
\usepackage{listings}
\usepackage{collcell}
\usepackage{multirow}
\usepackage[labelformat=parens]{subcaption}
\usepackage{url}
   \def\UrlFont{sf}

\usepackage{hyperref}
   \hypersetup{colorlinks=true, allcolors=DarkBlue}

\usepackage{pgfplots}
\usepackage{pgfplotstable}
   \pgfplotsset{compat=newest}
\usepackage{csvsimple}
\usepackage{tcolorbox}
\usepackage{enumitem}
\usepackage[ruled,noline,noend]{algorithm2e}
   
   \SetCommentSty{myalgocommentstyle}
   
\usepackage{afterpage}
\usepackage{longtable}
\usepackage[figuresright]{rotating}
\usepackage{pdflscape}

\newcommand*{\origrightarrow}{}

\renewcommand*{\textrightarrow}{\fontfamily{cmr}\selectfont\origrightarrow}

\newcommand{\FPtodo}[2][]{
  \def\FPtodoAuthorBGColor{orange}%
  \def\FPtodoAuthorFGColor{white}%
  \ifnum\pdfstrcmp{#1}{Mihai}=0.  \def\FPtodoAuthorBGColor{red}            \fi
  \ifnum\pdfstrcmp{#1}{Mike}=0    \def\FPtodoAuthorBGColor{green!60!black} \fi
  \textsf{\textbf{\fcolorbox{\FPtodoAuthorBGColor}{\FPtodoAuthorBGColor}{\textcolor{\FPtodoAuthorFGColor}{\textsc{#1}}}}{~\textcolor{\FPtodoAuthorBGColor}{#2}}}
}

\usepackage{fancyhdr}

\newcommand{\secref}[1]{\S\ref{#1}}

\title{Fingerprinting SDKs for Mobile Apps and Where to Find Them: \texorpdfstring{\\}\ Understanding the Market for Device Fingerprinting}
\author{Michael A. Specter}
\affiliation{\institution{Georgia Tech \& Google LLC}  \city{Atlanta, GA} \country{USA}}
\authornote{Email address: \textsf{\href{mailto:mobile-fingerprinting-sdks-paper@google.com}{mobile-fingerprinting-sdks-paper@google.com}} // \textsf{\href{mailto:specter@gatech.edu}{specter@gatech.edu}}}

\author{Mihai Christodorescu}
\orcid{0000-0001-5808-8015}
\affiliation{\institution{Google LLC} \city{Mountain View, CA} \country{USA}}

\author{Abbie Farr}
\affiliation{\institution{Google LLC} \city{Mountain View, CA} \country{USA}}

\author{Bo Ma}
\affiliation{\institution{Google LLC} \city{Mountain View, CA} \country{USA}}

\author{Robin Lassonde}
\affiliation{\institution{Google LLC} \city{Mountain View, CA} \country{USA}}

\author{Xiaoyang Xu}
\affiliation{\institution{Google LLC} \city{Mountain View, CA} \country{USA}}

\author{Xiang Pan}
\affiliation{\institution{Google LLC} \city{Mountain View, CA} \country{USA}}

\author{Fengguo Wei}
\affiliation{\institution{Google LLC} \city{Mountain View, CA} \country{USA}}

\author{Saswat Anand}
\affiliation{\institution{Google LLC} \city{Mountain View, CA} \country{USA}}

\author{Dave Kleidermacher}
\affiliation{\institution{Google LLC} \city{Mountain View, CA} \country{USA}}

\begin{document}
\def\UrlFont{\sf}

\pagestyle{plain}
\thispagestyle{empty}

\begin{abstract}

This paper presents a large-scale analysis of fingerprinting-like behavior in the mobile application ecosystem. 
We take a market-based approach, focusing on third-party tracking as enabled by applications' common use of third-party SDKs. 
Our dataset consists of over 228,000 SDKs from popular Maven repositories, 178,000 Android applications collected from the Google Play store, and our static analysis pipeline detects exfiltration of over 500 individual signals. 
To the best of our knowledge, this represents the largest-scale analysis of SDK behavior undertaken to date. 

We find that Ads SDKs (the ostensible focus of industry efforts such as Apple's App Tracking Transparency and Google's Privacy Sandbox) appear to be the source of only $30.56$\% of the fingerprinting behaviors. A surprising $23.92$\% originate from SDKs whose purpose was unknown or unclear. Furthermore, Security and Authentication SDKs are linked to only $11.7$\% of likely fingerprinting instances. 
These results suggest that addressing fingerprinting solely in specific market-segment contexts like advertising may offer incomplete benefit. Enforcing anti-fingerprinting policies is also complex, as we observe a sparse distribution of signals and APIs used by likely fingerprinting SDKs. For instance, only $2$\% of exfiltrated APIs are used by more than $75$\% of SDKs, making it difficult to rely on user permissions to control fingerprinting behavior.

\end{abstract}
\maketitle
\iffalse
{ \sffamily
\begin{tcolorbox}[colback=green!5!white,colframe=green!50!black,title=Conference Version]
A shorter version of this paper is published at the \href{https://www.sigsac.org/ccs/CCS2025/}{ACM Conference on Computer and Communications Security 2025 (CCS'25)}. The present version additionally includes:
\begin{itemize}[leftmargin=*]
    \item Descriptions of the static analyses performed,
    \item Description of the SDK-identification algorithm,
    \item Definition of the codebook used to categorized SDKs, and
    \item List of the APIs observed in fingerprinting-like behaviors.
\end{itemize}
\end{tcolorbox}
}
\fi

\section{Introduction}
\label{sec:intro}

\textit{Device fingerprinting} is a technique used to identify and track user devices by collecting a wide range of information about device-specific hardware, software, and configuration settings. 
The combination of these attributes creates a unique, or near-unique, digital ``fingerprint'' for that device.
This process has clear privacy concerns---fingerprinting identifiers can be collected without user control or notice, and persist over the device's lifetime regardless of most user privacy-seeking actions (e.g., clearing one's cookies, rotating advertising IDs, or enabling private browsing).

Both major mobile operating system vendors have undertaken significant efforts to limit the privacy impact of device fingerprinting. 
Apple has introduced policies that require apps to request user consent to collect tracking-relevant device data~\cite{ApplePrivacyPolicy}, and provide human readable explanations for the use of specific high entropy ``required reason'' APIs~\cite{AppleRequiredReason}.
Both Google and Apple's mobile platforms now require developers to provide nutrition label-style privacy information to the user~\cite{DescribingDataUseApple, googlePSL}, either as metadata submitted to their respective application stores or attached as part of the application itself.
Google is developing a privacy sandbox for the web~\cite{PrivacySandboxTechnologya}, and, on Android, a new sandbox that restricts third-party advertising libraries from accessing sensitive information available to the rest of the application~\cite{googleSDKRuntimeOverview}.
These interventions are promising---providing much needed transparency and accountability.

The success of such anti-fingerprinting efforts depend on the technical implementation, market fit, and intention of the application's developer. 
For example, Apple's anti-tracking and app transparency policies explicitly allow the collection of fingerprinting data for anti-fraud purposes~\cite{ApplePrivacyPolicy}, and Android's Privacy Sandbox focuses solely on isolating code from third-party advertisers. 
Apple's ``required-reason APIs'' approach also has limitations; it currently applies to just 30 APIs, and its effectiveness depends on what other data points are collected across the wider fingerprinting ecosystem. 
Therefore, characterizing the technical implementation of fingerprinting in the wild and understanding stakeholders involved would provide invaluable insight into the effectiveness of these interventions.

This paper presents a comprehensive, large-scale analysis of device fingerprinting practices within the Android application ecosystem. 
We adopt an empirical approach centered on the identification of  third-party Software Development Kits (SDKs) integrated in mobile applications, measuring their market reach, tracking methodologies, and privacy impact. 
To the best of our knowledge, this research represents the most comprehensive analysis of SDK behavior regarding privacy-invasive practices like device fingerprinting, with an extensive dataset of over 228,000 unique SDKs and 178,000 Android applications. 
Our methodology attempts to enable a more nuanced understanding of the scale and scope of device fingerprinting in mobile ecosystems; we avoid applying our own potentially biased or narrow definitions on fingerprinting behavior, and adopt a number of techniques to provide reliability and consistency when subjective analysis is unavoidable.

While many studies have measured the impact of fingerprinting (see \secref{sec:related} for related work), there is a dearth of knowledge surrounding the purpose of fingerprinting behavior. 
For example, no prior study has attempted to understand what kinds of third parties collect sufficient information to fingerprint a device, and how this fits in with the needs of the first party developer. 
Characterizing the overall problem from the perspective of developers can help determine why these techniques are used, provide invaluable insight into what is required to better preserve user privacy, and interpret the value of assumptions underlying current and proposed enforcement methods.

There are a number of challenges that significantly complicate our study. 
Any fingerprinting-detection mechanism will likely be incomplete, as there are many (potentially stealthy) methods of collecting entropy from a device, including timing information, instruction execution quirks, and other hardware-specific sources. 
Categorization and analysis of SDKs is also a difficult task---while \textit{applications} self-label their use and market-fit, current SDK distribution methods do not require SDK authors to provide significant descriptions of their code. 
We describe the solutions to these and other challenges in depth in \secref{sec:method}.

One important challenge is definitional: The claim that a service is fingerprinting suggests intent of the author of the code, which is most often practically unknowable.
Applications may collect sufficient information to uniquely identify a device for any number of reasons, including analytics, crash reporting, anti-fraud, or through normal operation of the application itself. 
We emphasize that our study is purely observational---we measure \textit{fingerprinting behavior}, and ascribe no motivation to the authors of the code. We also emphasize that our choice to examine the Android ecosystem is entirely due to convenience, and that our results are likely to extend to iOS as well. As noted in prior work~\cite{kollnigAreIPhonesReally2022}, whereas Android's open ecosystem allows for scalable analysis, iOS's digital rights management scheme actively hinders the same. 

We answer the following research questions:
\begin{description}
\item[RQ1:] What types of behaviors do \textit{self-identifying} fingerprinting SDKs exhibit?
\item[RQ2:]What are the stated purposes of SDKs with likely fingerprinting behavior?
\item[RQ3:] What kinds of apps use SDKs with likely fingerprinting behavior, and how prevalent are these SDKs in real world apps?
\end{description}

We find that many kinds of SDKs collect sufficient information to track a user (at least $20$ signals exfiltrated per SDK), and that there is a large diversity in signals collected (SDKs exfiltrate $75.5$ signals on average, out of a total of $504$ unique signals observed across the SDK dataset).
Though ads do make up a significant portion ($\approx 30$\%) of the SDKs that exhibit fingerprinting behavior, a surprising number of fingerprinting-like SDKs used in common Android applications have unclear functionality and lack significant description for categorization ($\approx 24$\%).
Anti-fraud and analytics services were also prevalent in our dataset, indicating that more research must be done to create privacy-preserving alternatives to fingerprinting as used in such functionality. 
Finally, SDKs that exhibit likely fingerprinting behavior are disproportionately popular---roughly $10\times$ more installs than non-fingerprinting alternatives---and individual SDKs are likely to exist across multiple application market segments (e.g., health and dating).

\iffalse
\subsubsection*{Roadmap}  We begin in \secref{sec:background} with important background and prior work, as well as a short overview of the fingerprinting threat model. In \secref{sec:method} we describe our dataset, analysis pipeline, and labeling methodology. Next, \secref{sec:results} presents an overview of our results, and we conclude with a discussion and examination of our study's limitations in \secref{sec:discussion}.
\fi

\newcommand{\circNum}[1]{\textcircled{\raisebox{-0.5pt}{\sffamily\small #1}}}

\begin{figure*}[t]
    \centering
    \Description{This image is a flowchart illustrating a multi-step process related to identifying and classifying Software Development Kits (SDKs) within applications. The diagram uses rectangular boxes for process steps and cylinder shapes for databases, connected by arrows indicating the flow of data or action. Here's a breakdown of the elements. Process Steps (Rectangular Boxes with Numbers). There are five rectangular boxes, numbered sequentially, representing stages in the process: Dataset Collection: Located at the far left. Text below it describes fetching datasets of SDKs and applications from Maven Repos and Google Play. A small Maven and Google Play logo are also present near this box. Seed Set & Manual Signal Extraction: Located to the right of step 1. Text below it explains that manual reverse engineering is performed to collect a set of signals, yielding a "Seed Set". Automated Signal Exfiltration Detection: Located to the right of step 2. Text below it states that static taint analysis determines which signals are exfiltrated. A small label "Fingerprinting Signals" is on the arrow pointing to this box from step 2. SDK Labeling: Located to the right of step 3. Text below it indicates that SDKs are manually labeled by type and kind. App-SDK Matching: Located at the far right. Text below it explains that static analysis confirms if an application contains a specific SDK. Databases (Cylinder Shapes): There are three cylinder shapes representing databases: App DB: Located above steps 1 and 5. SDK DB: Located above steps 1, 2, and 3. Extended Set SDKs: Located above steps 3 and 4. Connections (Arrows): Arrows connect the process steps and databases, showing the direction of flow. Some arrows are thicker than others, potentially indicating a more significant or primary flow. Arrows from "Dataset Collection" (1) point to "App DB" and "SDK DB". Thick arrows connect "SDK DB" to "Seed Set & Manual Signal Extraction" (2) and "Automated Signal Exfiltration Detection" (3). An arrow labeled "Fingerprinting Signals" connects "Seed Set & Manual Signal Extraction" (2) to "Automated Signal Exfiltration Detection" (3). A thick arrow connects "Automated Signal Exfiltration Detection" (3) to "Extended Set SDKs". An arrow labeled ">= 20 Signals" points from the arrow entering "Extended Set SDKs" from step 3, suggesting a condition for adding to this database. Arrows connect "Extended Set SDKs" to "SDK Labeling" (4) and "App-SDK Matching" (5). A thick curved arrow connects "App DB" to "App-SDK Matching" (5). In summary, this flowchart outlines a process for identifying and characterizing SDKs. It starts with data collection, moves to manual and automated analysis steps, populates and utilizes databases (App DB, SDK DB, Extended Set SDKs), involves manual labeling, and concludes with matching apps to SDKs. The arrows and labels provide context on the flow of information and dependencies between the steps.}
    %\includesvg[width=\textwidth]{figures/overview.svg}
    %\includegraphics[trim={0 .40cm 0 0}, clip, width=\textwidth]{figures/app_sdk_probabilities.pdf}
    \includegraphics[width=\textwidth]{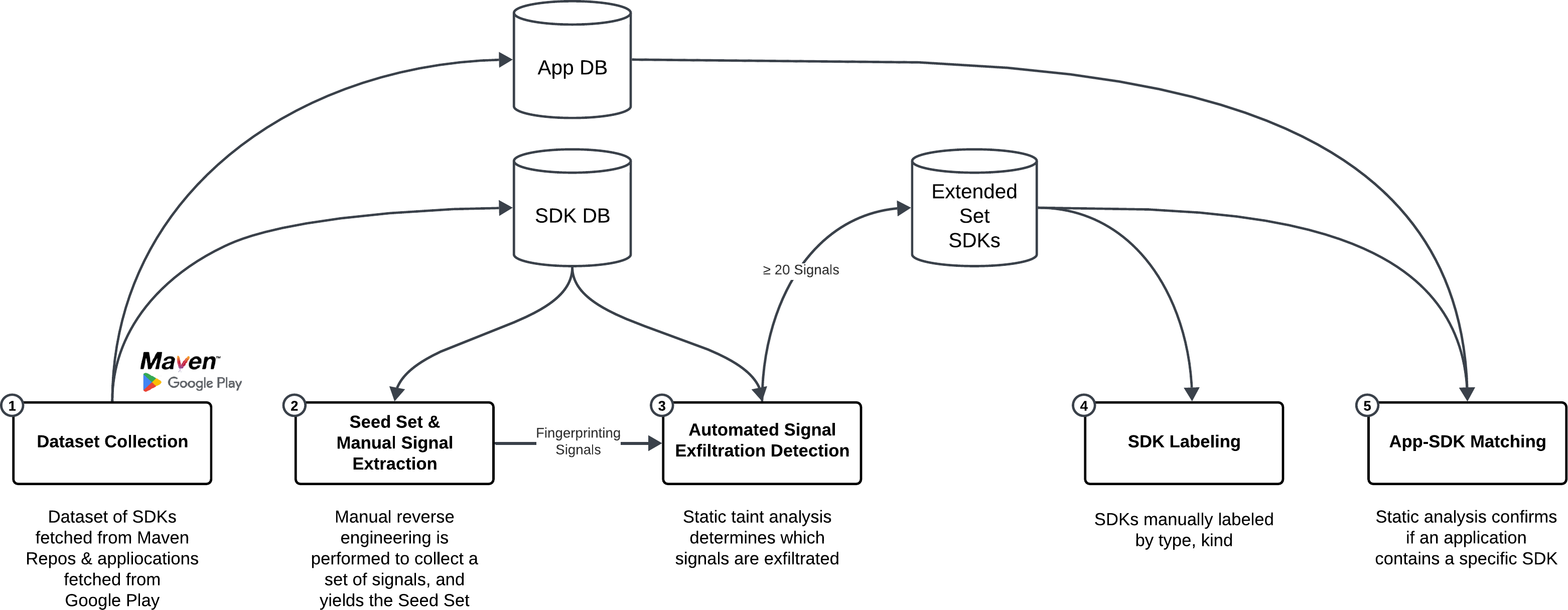}
    \caption{
An overview of our analysis pipeline. We begin by \circNum{1}~fetching apps, SDKs, and associated metadata from a series of Maven repositories and the Google Play app store, only selecting applications installed on $>10k$ active devices (\secref{sec:meth_data}). We continue by \circNum{2}~extracting a seed set of SDKs whose copy indicates that they are fingerprinting (\secref{sec:meth_Seed}). We then \circNum{3}~use static taint analysis to determine which SDKs exfiltrate these signals (\secref{sec:meth_automated_signal}), and \circNum{4}~manually label the resulting SDKs to determine their market fit (\secref{sec:meth_sdk-labeling}). Finally, we \circNum{5}~perform another round of static analysis to determine which applications contain which SDKs (\secref{sec:meth_app-sdk-matching}).
    }
    \label{fig:overview}
\end{figure*}

\section{Background \& Related Work}
\label{sec:background} \label{sec:related}

To the best of our knowledge, our work is the first large-scale study of native app-based device fingerprinting in the wild.\footnote{We make a distinction here between \textit{web-based} and \textit{on-device} app fingerprinting. E.g., prior studies have looked into the prevalence of fingerprinting on the web. Here we are exclusively interested in fingerprinting that occurs on-device outside of a web context.} No prior research has attempted to understand why this phenomenon is common, or the market surrounding the use of these tools. Much of the existing literature stems from examining fingerprinting as an attack, with a focus on novel methods of fingerprinting.

\subsubsection*{Android applications \& SDKs}
Android applications may be written in any language, and can be installed from arbitrary sources including the Play store, secondary app stores, side-loading, or may come pre-loaded on-device from the manufacturer. As a result, much of Android's security model revolves around sandboxing applications using a combination of SELinux's SEPolicy and standard Linux UID-style access control mechanisms. In addition to sandboxing, access to certain sensitive data is declared via metadata provided by the app, and enforced via both install and run-time permissions checks~\cite{mayrhofer_android_2021}. 
Unless manually sandboxed, Android third party libraries (called \textit{SDKs}) execute in the first-party application context, and therefore enjoy the same permissions as the first-party application.

SDKs may be distributed as either raw code or automatically downloaded at build time from any number of repositories or build systems. 
A commonly used build system is the \textit{Maven} format, an open standard for Java dependency resolution. 
In practice, distribution of SDKs is commonly done using a build tool called Gradle, which loads SDKs from any number of public Maven repositories.

\subsubsection*{Fingerprinting signals}
In this paper we define a \emph{signal} to be an individual data point collected from a device. Different papers have different terms for a unique datapoint in a fingerprint, Eckersley \cite{10.1007/978-3-642-14527-8_1} calls it a variable. Signals may be obtained from many sources including API calls, common files on the platform, system properties, hardware quirks, or runtime environment values.

\subsubsection*{Diversity of signals}
Though fingerprinting is commonly discussed in the context of browsers~\cite{englehardt_online_2016,10.1007/978-3-642-14527-8_1}, there is a rich literature surrounding the many methods of fingerprinting via native code. 
Fingerprintable components include the microphones and speakers~\cite{10.1145/2660267.2660325,10.1145/2660267.2660300,bojinov2014mobile}, the accelerometer, the gyroscope, and the magnetometer~\cite{9265280,8835276,9628005,conf/ndss/DeyRXCN14,bojinov2014mobile,8809830,10.1007/978-3-319-30806-7_7}, the hardware clock~\cite{1453529,10.1145/3243734.3243796}, the camera~\cite{10.1007/978-3-319-25512-5_12,10.1007/978-3-030-29959-0_22}, the GPU clock~\cite{nakibly2015hardware}, and the battery~\cite{8368345,10.1007/978-3-319-29883-2_18}. Fingerprinting of the system or the apps ranges from specific APIs~\cite{10.1145/3407023.3407055}, to system configuration (e.g., via \texttt{procfs}~\cite{10.1145/3196494.3196510,255288}), user settings~\cite{kurtz2016fingerprinting}, and browser configuration~\cite{10.1007/978-3-642-14527-8_1,7958618}. 
Fingerprinting can also be done by communicating with neighboring devices via short-range radio protocols such as Bluetooth~\cite{10.1145/3176258.3176313}. Distinct signals can be combined to increase accuracy~\cite{7934347,Cao2017CrossBrowserFV}.

\subsubsection*{Detection \& prevention}  
Detection and prevention methods include ML classification of API calls~\cite{bahrami2021fpradar, 10.1007/978-3-319-20810-7_21, 10.1145/3590777.3590790}, re-calibrating sensors~\cite{das2015exploring}, changing system settings~\cite{10.1145/2818000.2818032}, adding random noise to collected data~\cite{10.1145/2736277.2741090, das2015exploring}, or using taint tracking to identify exfiltration of fingerprintable data~\cite{10.1007/978-981-16-9229-1_4}.
Permissions systems do not offer adequate protection against fingerprinting~\cite{ 7738449, 10.1145/3360664.3360671}.

\subsubsection*{Prior measurement studies} 
There have been a number of studies that measure the use of fingerprinting, though the majority focus on the web~\cite{englehardt_online_2016,nikiforakisCookielessMonsterExploring2013}.
A significant challenge appears to be the ever-growing surface of APIs, which have been quickly adopted by fingerprinters~\cite{bahrami2021fpradar}. 

Existing analyses of Android native systems are comparatively rare. 
Longitudinal analyses not only highlighted this distinctive, mobile-specific flavor of SDK-based tracking in general, but also showed that the privacy risk across time and app versions varies greatly with little correlation to existing enforcement approaches~\cite{Ren2018BugFI}. 
Han et. al.~\cite{osti_10187034} find that the presence of privacy-risky behavior (including fingerprinting) does not seem to vary by the cost of the app, with free apps and paid apps sharing similar sets of third-party SDKs or dangerous permissions.

The closest study to ours is Torres et al's 2018 work \cite{ferreira2018investigating} on identifying fingerprinting in applications. %
They find that fingerprinters on mobile devices rely more on categorical signals and less on side channels, and argue that detection and prevention of fingerprinting on mobile is distinct from web browsers. 
Our work uses a significantly increased scale in terms of signals, SDKs, and applications (30k vs our 178k), presenting a more complete understanding of the ecosystem, in addtion to SDK labels and further statistics.

\section{Methodology}

\label{sec:method}

In this section, we provide an in-depth discussion of our analysis pipeline. 
We depict our overall process in Figure~\ref{fig:overview}, and outline a summary below:

\begin{enumerate}
    \item \textbf{Dataset Collection (\secref{sec:meth_data}):} We fetch a dataset of SDKs by crawling popular Maven repositories and a dataset of applications from the Google Play Store.
    
    \item \textbf{Seed Set \& Manual Signal Extraction  (\secref{sec:meth_Seed}): } We use the advertising copy published by each individual SDK to generate a \textit{Seed Set} that in public statements self-announce as fingerprinters. We then manually reverse engineer these SDKs to extract what \textit{signals} each exfiltrate.
    
    \item \textbf{Automated Signal Exfiltration Detection (\secref{sec:meth_automated_signal}):} Using static taint analysis in tandem with data generated from our Seed Set, we determine if an SDK exfiltrates relevant fingerprinting signals. We call all SDKs that perform enough exfiltration to be exhibiting fingerprinting behavior the \textit{Extended Set}.
    
    \item \textbf{SDK Labeling \& Analysis (\secref{sec:meth_sdk-labeling}):} Unlike applications, SDKs are unlabeled, and do not carry metadata associated to market, use-case, or intended audience. To provide adequate statistics and market information, we manually label all SDKs in the Extended Set. To avoid bias and meaningless labels, we borrow coding techniques from the HCI community, iteratively developing a codebook and reaching consensus on SDK label definitions and assignments. 
    
    \item \textbf{App-SDK Matching  (\secref{sec:meth_app-sdk-matching}):} To provide statistics on the use of each SDK, we must first determine which SDK exists in which application. We accomplish this through a series of static analysis techniques.
\end{enumerate}

There are a number of reasons we focus on SDKs rather than applications as a whole. One may expect an SDK to be self-contained and have explicitly publicized functionalities. 
As with any modern development environment, SDKs use in applications is incredibly common, with the majority of apps using third-party libraries to support a variety of core functions.
Finally, an emphasis on SDKs also makes the potential harm from third parties far clearer: users are more likely to understand and trust an application they have actively installed, but may be unaware of the transitive trust they have placed on the third party SDKs and services used by an app.

\subsection{Dataset Collection}
\label{sec:meth_data}

\subsubsection*{Application Dataset} 
\label{sec:meth_app_data}
We collected 3,025,417 APKs\footnote{APK is the file format of Android apps. In the rest of the paper we use the term ``APK'' as shorthand for one Android app.} published on the Google Play store over almost 18 months (from January 2023 to May 2024). 
We supplement this set of APKs with each application's \textit{total audience size} -- the number of active devices that an individual APK has been installed on. An active device is a device that has been turned on at least once in the previous 30 days~\cite{googleViewAppStatistics}.

To avoid biasing our sample set with applications that lack a significant user base, we limit our analysis to applications active on the Google Play store with a total audience size of over 10,000 from April 13, 2024 to May 13, 2024. In total, this covers 178,054 applications.
While we cannot estimate how often these applications were launched, the audience-size metric assures that devices on which the apps were installed were in active use.
We approximate the market reach of an app by summing all active 30-day installs of these apps --- we note that this may double-count users who install, remove, and then reinstall the same app, as well as installs by the same user on one device under multiple profiles (e.g., personal and work) or on multiple devices (e.g., phone and tablet).

\subsubsection*{SDK Dataset} 
There is no single source of information for SDKs, instead developers supply their build system with a URL of a particular Maven repository and library ID. We collate our dataset using a custom crawler, extracting all SDKs from 9 separate large-scale Maven repositories: JCenter, Maven Central, Google, Sonatype, Spring.io, Jitpack, Bintray, and Artifactory. 
From these repositories, we fetched a dataset of 228,598 SDKs as well as each SDK's associated metadata. 
We excluded all SDKs whose version label included one of the words \{``alpha'', ``beta'', ``test'', ``dev'', ``debug'', ``qa''\}.

\subsection{Seed Set and Manual Signal Extraction}
\label{sec:meth_Seed}

From our dataset of SDKs we select a \textit{Seed Set} of SDKs that, in their advertising copy or other metadata, openly admit to collecting information for the purpose of fingerprinting. 
We then manually reverse engineered each SDK, confirmed that the SDK was collecting a nontrivial set of device data, and extracted a list of signals that the SDK uploaded to a server. 
To avoid mislabeling, each candidate SDK is then reverse engineered \textit{again} by a second analyst, who independently confirms the list of signals. 
In total, our Seed Set contains 14 SDKs, reporting over 500 distinct signals. The results from this effort are reported in more detail in \secref{sec:rq1}.

\subsection{Automated Signal Exfiltration Detection}
\label{sec:meth_automated_signal}

We collected an \textit{Extended Set} of \textit{fingerprinting-like SDKs}, consisting of SDKs that are similar in terms of collected data to the Seed Set. 
To obtain this set, we developed a static analysis suite that performs information-flow analysis on SDKs and their dependencies, and selected SDKs with sufficient signal overlap with our Seed Set. 
Note that SDKs in the Extended Set may collect signals beyond those found in the Seed Set.

To avoid over-claiming the existence of fingerprinting behavior in our dataset, we only include an SDK if it exfiltrates \textit{more than the lowest number of signals collected by any SDK in our Seed Set.} 
Put another way, we only consider an SDK to be exhibiting fingerprinting behavior if it exfiltrates more than what is uploaded by an SDK that openly admits to fingerprinting.
Note that this is a conservative estimate---it is likely that more sophisticated estimates of entropy would indicate that an SDK could uniquely identify a user using \textit{less} information than what we examine here. 
The tradeoff here is intentional; our goal is to provide upper-bar estimates without indulging in more complicated analyses.
This step results in 723 distinct SDK families, each with multiple versions, for a total of 14,178 SDK versions.

We built a static-analysis suite for Android APK and SDK analysis,
and deployed an interprocedural, context-, field-, object-sensitive taint-flow tracking algorithm
for fingerprinting detection.
It works by tainting all fingerprinting-related data with meta information
and then propagating taint at the instruction level, so that the transparency of
the flow information can be achieved, allowing us to reconstruct the taint flow path
to independently verify exfiltration.
We claim no novelty for this analysis, though some implementation details may be of independent interest, 
so we include these in Appendix~\ref{sec:det-details}.
 
\begin{table}
   \caption{SDK Label Definitions. Each SDK receives one label based on its Maven metadata and website description, not based on its code. See Table \ref{tab:codebook_full} in Appendix \ref{sec:further_defs} for more formal descriptions used in our labeling process.}
   \label{tab:purposes}

   \centering
   \begin{tabularx}{\columnwidth}{lX}
      \toprule
      \textit{\begin{tabular}{@{}l@{}}SDK Label \end{tabular}}  & \textit{Description} \\
      \midrule \arrayrulecolor{lightgray}
      Advertising & {Supports displaying ads, ads bidding, ads targeting, ads mediation, or analytics for the purpose of monetization or conversion (Ex: AppLovin, Teads)} \\
      \cmidrule{1-2}
      Analytics & {Monitors and reports on app health (examples: TOAST Logger, RichAPM Agent), or collects the user's behavior in app (Ex: Pushwoosh, Acoustic Tealeaf)} \\
      \cmidrule{1-2}
      \parbox[t]{2cm}{Security \& \\  Authentication}  & {Implements user authentication (E: Passbase, Ondato), detects fraud and related security anomalies (Ex: Incognia, SEON), and payment functionality (Ex: Alipay, PayPal).} \\
      \cmidrule{1-2}
      Tools / Other & {Provides navigation (Ex: Tencent Map Nav, Radar), object or person tracking (Ex: BeaconsInSpace, Foursquare Movement), communication with social networks (Ex: Facebook, Chat SDK), or other well-defined functionality (Ex: iZooto App Push, GameUp)} \\
      \cmidrule{1-2}
      \parbox[t]{2cm}{Unclear / \\ Not Found }& {Purpose or functionality could not be determined from online metadata} \\
      \arrayrulecolor{black} \bottomrule
   \end{tabularx}
\end{table}

\subsection{SDK Labeling}
\label{sec:meth_sdk-labeling}

While developers provide an attestation of the categorical use-case of their \textit{application} as a part of submission to the Google Play store, there is no equivalent process for \textit{SDKs}. 
Indeed, Maven repositories usually provide only the name of the SDK, a short explanation, and a link back to the originator of the code. This metadata is often incomplete, further obfuscating the use of the SDK.
Other datasets are more complete, including the Google Play SDK Index~\cite{googleSDKIndex}, but provide only a small database of SDKs.

Here we borrow techniques from the HCI community and treat the problem as a manual labeling task. 
Label definitions were collaboratively developed by a team of five expert coders based on the metadata of the SDK, including the SDK's description in Maven and the content of its developer's website, from a sample of 100 random SDKs in our dataset. For brevity, an informal description of these categories is in Table~\ref{tab:purposes}, and a full explanation (including sub-categories) is in Table~\ref{tab:codebook_full}.
Upon reaching saturation, we split the remaining SDKs between reviewers such that each SDK was independently examined twice.

For efficiency, we limited our labeling effort to the 723 SDK families that have been detected in our application dataset, under the assumption that all versions of the same SDK have an equivalent use case and should share the same label.
The resultant definitions were robust, with reviewers usually agreeing on SDK labels; the Krippendorff's alpha inter-rater reliability score of the independent labeling step was 0.804. 
Finally, all label disagreements were resolved and re-labeled in a meeting of the full group, meaning that any labeling disagreement was addressed by comparing labels from all five coders. %

\subsection{App--SDK Matching}
\label{sec:meth_app-sdk-matching}

Inspired by the large body of work in SDK identification for Android apps~\cite{9286020, 9551847, 9542854, 10.1145/2976749.2978333,  10.1145/2889160.2889178, 7985674, 8543426, 7985674, 10.5555/3288994.3289051, 10.1109/ICSE43902.2021.00150,9120346}, we created an SDK-identification pipeline using a fine-grained code similarity metric that can be aggregated across code units (e.g., classes, modules) and packaging units (e.g., SDKs, SDK versions). The similarity metric relies on identifiers for system APIs (e.g., operating system calls, standard library calls), opcode frequencies, framework APIs, and string constants. 

In our design, we choose parameters for this similarity metric, including the percentage of APK code similar to a known SDK sufficient to declare the SDK as present in the APK, to ensure that our results limited false positives at the risk of some false negatives. In other words, we may miss the presence of an SDK in an APK, and thus the statistical analysis in the rest of the paper provides lower bounds for the prevalence of fingerprinting SDKs. We include a detailed description of our approach in Appendix~\ref{sec:sdk-ident-details}.

\section{Results}
\label{sec:results}

We analyzed the Seed Set, Extended Set, and their prevalence in our application dataset to answer our three research questions (\secref{sec:intro}).

\subsection{RQ1: Self-Identified Fingerprinters}
\label{sec:rq1}

\begin{figure*}
    \centering
    \Description{This image contains two linked plots displayed one above the other, along with a vertical axis label on the left that applies to the entire figure. Overall Vertical Axis: The vertical axis on the left is labeled "API prevalence". Top Plot: Chart Type: Vertical bar chart. Horizontal Axis (Implicit): Represents different API IDs, arranged sequentially from left to right. There are many bars, indicating a large number of API IDs. The tick marks are not labeled numerically. Vertical Axis: Labeled "API prevalence". The scale ranges from 0.00 to 1.00, with tick marks at 0.00, 0.25, 0.50, 0.75, and 1.00. Bars: Blue vertical bars rise from the horizontal axis. Each bar corresponds to an API ID, and its height represents the API prevalence (percentage of Seed Set SDKs that exfiltrate that API). Trend: The heights of the bars vary significantly across the different API IDs, suggesting a wide range in how prevalent different APIs are among the Seed Set SDKs. There are some bars that reach close to the top of the scale (1.00), but many are in the lower half. Bottom Plot: Chart Type: Swarm plot or scatter plot with points aligned vertically for each category. Horizontal Axis: Labeled "API id", aligning with the horizontal axis of the top bar chart. Again, individual API IDs are represented horizontally, although not explicitly labeled. Vertical Axis: Labeled with the names of different SDKs: Forter, Fingerprint.js, Accertify, Microsoft Dynamics, ThreatMetrix, Shield, Castle, TransUnion TruValidate, incognia, Socure, Seon, Ravelin, Kaspersky AntiVirus SDK. Data Points: Colored dots are plotted horizontally for each SDK. Each dot represents an API that is exfiltrated by that specific SDK. The color of the dots is consistent for each SDK row (e.g., Forter has pink dots, Fingerprint.js has orange dots). Relationship to Top Plot: The horizontal position of the dots in the bottom plot aligns with the corresponding API ID bars in the top plot. This means that for a given API ID (a vertical slice of the chart), if an SDK has a dot at that position, it exfiltrates that API. The height of the bar above shows the overall prevalence of that API across all the SDKs listed. In summary, this figure presents a combined visualization. The top bar chart shows the overall prevalence of different APIs across a set of SDKs, while the bottom plot shows which specific SDKs exfiltrate each API, providing a detailed view of which APIs are collected by which fingerprinting SDKs in the Seed Set.}
    \includegraphics[width=\textwidth]{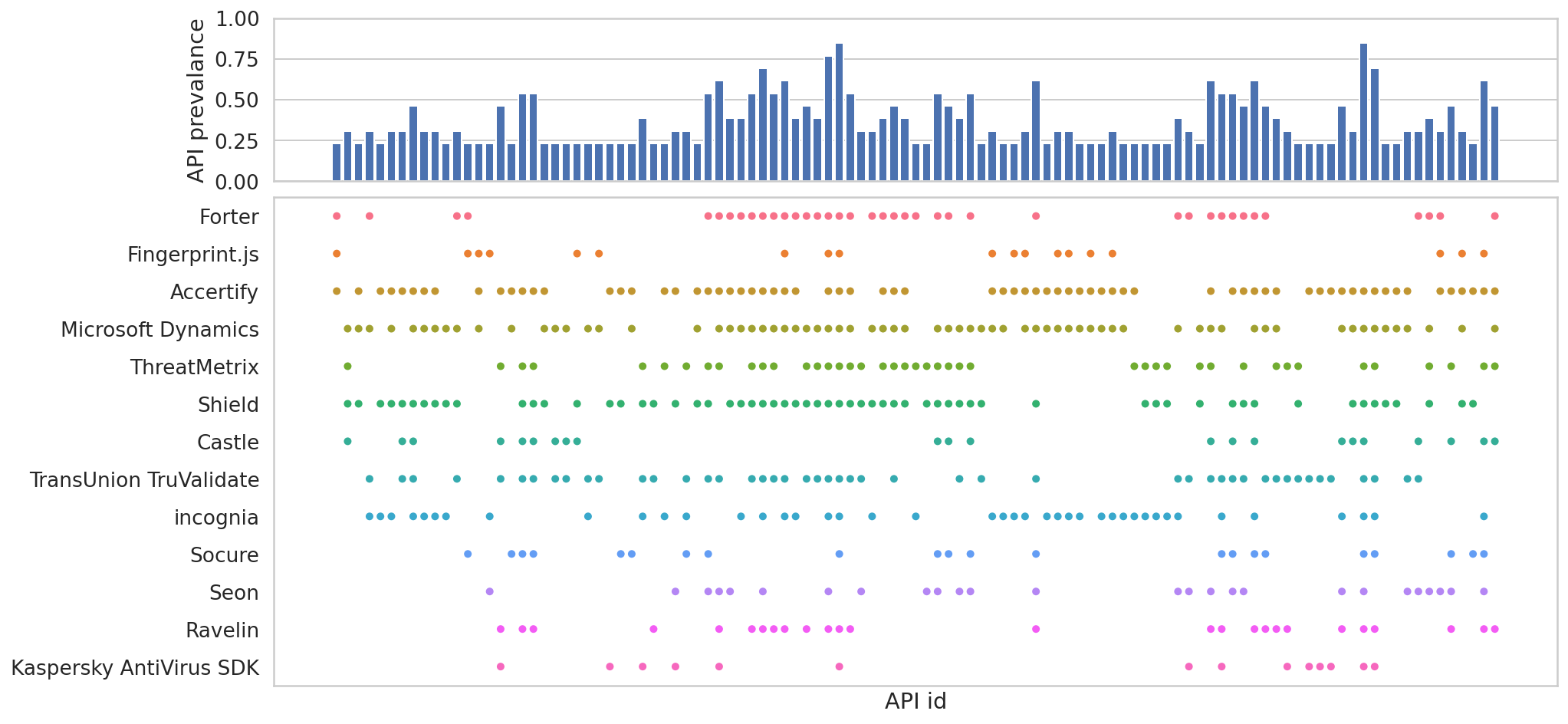}
    \caption{Map of signals collected by known fingerprinting SDKs in the Seed Set, with one dot per API exfiltrated. The top plot shows the percentage of Seed Set SDKs that exfiltrate that API; 80\% of APIs are exfiltrated by fewer than 50\% of SDKs, and only 2\% of APIs are exfiltrated by 75\% of SDKs.}
    \label{fig:heatmap_of_sdk_collect}
\end{figure*}

\begin{table}[tbp]
   \caption{Seed Set SDKs. List of SDKs that openly admit to fingerprinting, through advertising copy, developer documentation, or other copy. The number of raw number of signals they collect is on the right. Unique Signals is not a total, but the set union of all signals without repetition.}
   \label{tab:knownfp}

   \centering
   \begin{tabular}{lr}
      \toprule
      \textit{Name} %
                    & \textit{Signals} \\ %
        \midrule
        Seon                      \hfill & 43 \\
        Forter                     & 69 \\
        Kaspersky AntiVirus SDK    & 20 \\
        Accertify (InAuth)         & 213 \\
        Castle                     & 31 \\
        Microsoft Dynamics 365     & 128  \\
        IP Quality Score           & 58 \\
        Fingerprint.js             & 30 \\
        Shield                     & 148 \\
        ThreatMetrix (Lexus Nexus) & 94 \\
        Ravelin                    & 30 \\
        TransUnion TruValidate     & 55 \\
        Socure                     & 43 \\
        Incognia                   & 81 \\
        \midrule
        Unique Signals  &  504 \\
        \bottomrule
   \end{tabular}
\end{table}

\begin{quote}
    \textit{What types of behaviors do \textit{self-identifying} fingerprinting SDKs exhibit?}
\end{quote}

We find 14 different SDKs that admit to fingerprinting in the wild, listed in Table \ref{tab:knownfp}. Our manual analysis found that  fingerprinting libraries exfiltrate a minimum of 20 unique signals, and an average of 75.5. 
Notably, the techniques used by these SDKs were straightforward, with no SDK attempting to collect more than what was available from framework-level API calls.
This is a departure from what has been previously measured in the web context (e.g. audio context fingerprinting~\cite{englehardt_online_2016}), as well as the more advanced, hardware-focused techniques discussed by the academic community.

\begin{figure}
   \Description{This image is a heatmap illustrating the pairwise similarity between different mobile SDKs. The data is presented as a matrix, with each row and column representing one of the tools. The color of each cell indicates the level of similarity between the tools in that row and column, with a color bar on the right showing the mapping from color to similarity value (ranging from 0 to 1). The similarity values are also displayed numerically within each cell. Here's a breakdown of the visual elements. Layout: The heatmap is a square grid. Axes: The vertical axis and horizontal axis are labeled with the names of the fraud prevention tools. These names appear to be: Microsoft Dynamics, ThreatMetrix, Shield, Seon, Ravelin, Forter, TransUnion TruValidate, Socure, Incognia, Fingerprint.js, Castle, Accertify, Kaspersky AntiVirus SDK. Color Scale: A color bar is positioned to the right of the heatmap. It displays a gradient of colors, transitioning from light blue/gray at the bottom (representing lower similarity) to red at the top (representing higher similarity). Tick marks and numerical labels on the color bar indicate the correspondence between color and the similarity value (ranging from 0.0 to 1.0). Cell Colors and Values: Each cell in the grid is colored according to the similarity value between the corresponding row and column tools. The numerical similarity value, formatted to two decimal places, is also displayed within each cell in white text. Diagonal: The diagonal cells of the matrix are colored red and have a value of 1.00. This represents the similarity of each tool to itself, which is expected to be perfect. In summary, the heatmap provides a visual representation of the relationships between these fraud prevention tools, allowing viewers to quickly identify which tools are more or less similar to each other based on the color intensity of the corresponding cells.}
   \includegraphics[width=\columnwidth]{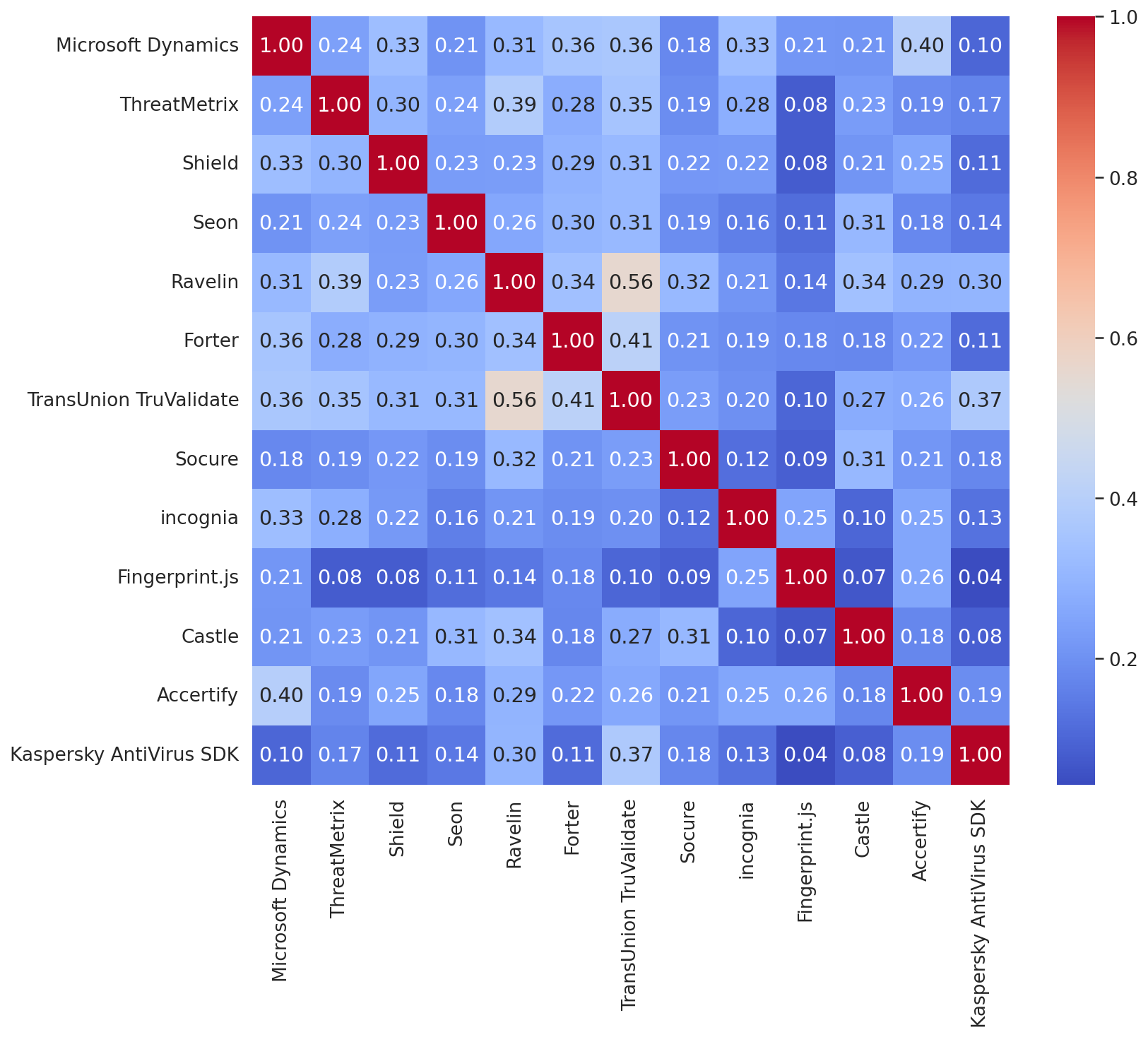}
   \caption{Cosine similarity between Seed Set SDKs, each represented as a one-hot encoding vector of their specific APIs used.}
   \label{fig:cosine}
\end{figure}

We examine the distribution of signals collected by SDKs in the Seed Set, and present an overview in Figure~\ref{fig:heatmap_of_sdk_collect}. 
Though the total number of signals collected is 1043, there are only 504 \textit{unique} signals, with less than half of all signals collected by at least two SDKs.
The set of fingerprinting signals collected are relatively sparse, as the individual signals that SDKs choose to select are somewhat dissimilar---\autoref{fig:cosine} displays the cosine similarity between different fingerprinting SDKs, with only two SDKs scoring above $0.5$ (Transunion and Ravelin). Of the $504$ unique signals collected, we find that only $21$ individual APIs were collected by more than half of the SDKs in our Seed Set, by a maximum of 11 SDKs. We conclude that cosine similarity of individual signals (as used in prior work~\cite{ferreira2018investigating}) is unlikely to be an effective detection mechanism. 

Some SDKs stand out with unique API usage patterns. For instance, Forter, Accertify, Microsoft Dynamics, and Shield appear to use a much broader range of APIs compared to others. This could suggest that these SDKs are more complex or serve a wider range of functionalities. Conversely, Kaspersky AntiVirus SDK appears to use a very limited set of APIs, possibly reflecting its more focused security purpose. Under the assumption that all of these SDKs perform fingerprinting equally well, such breadth of API usage patterns implies that some APIs provide more valuable (i.e., more fingerprintable) signals than others.

\subsection{RQ2: Purposes of Likely Fingerprinters}
\label{sec:rq2}
\begin{quote}
    \textit{What are the stated purposes of SDKs with likely fingerprinting behavior?}
\end{quote}

\begin{figure}
   \pgfplotstableread[col sep=comma, trim cells, header=true]{data/sdk-labels.data}\sdkLabelsTable
   \pgfplotstablesort[sort key={Count},sort cmp={int <}]\sortedSDKLabelsTable\sdkLabelsTable
   \definecolor{DodgerBlue}{rgb}{0.12, 0.56, 1.0} % Workaround for dumb error
   \centering
   \Description{This is a horizontal bar chart showing the frequency of different categories, ordered from highest to lowest frequency. Each bar represents a category, and its length indicates the number of items in that category. The categories are listed vertically on the left side of the chart, and the corresponding number is shown to the right of each bar. Here's a breakdown of the elements. Chart Type: Horizontal bar chart. Categories: The categories are labeled on the left vertical axis and are: Ads, Unclear / Not found, Tools / Other, Security and Authentication, Analytics. Bars: Each category has a blue horizontal bar extending to the right. The length of the bar corresponds to the numerical value associated with that category. Values: To the right of each bar, the specific count for that category is displayed as a number: Ads: 221, Unclear / Not found: 173, Tools / Other: 167, Security and Authentication: 85, Analytics: 77. Order: The bars are arranged from the top downwards, representing the categories in descending order of their values, from the highest (Ads) to the lowest (Analytics). Axis: There appears to be a faint vertical line on the left acting as the baseline for the bars. In essence, the chart visually communicates the relative frequency of items within each of the listed categories, making it easy to see which categories are most and least represented.}
   \begin{tikzpicture}
      \begin{axis}[
            width=0.70\columnwidth,
            xbar,
            ytick=data,
            axis y line=center,
            y axis line style={-},
            enlarge y limits=0.20,
            yticklabels from table={\sortedSDKLabelsTable}{SDK Category},
            y tick label style={anchor=east,font=\footnotesize\sffamily},
            x axis line style={draw=none},
            axis x line=none,
            nodes near coords,
            every node near coord/.append style={font=\footnotesize\sffamily,color=black},
         ]
         \addplot+ [draw=blue,fill=DodgerBlue] table [y expr=\coordindex] {\sortedSDKLabelsTable};
      \end{axis}
   \end{tikzpicture}
   \caption{Prevalence of purposes across the Extended Set of 723 likely fingerprinting SDKs.} 
   \label{fig:purpose-stats}
\end{figure}
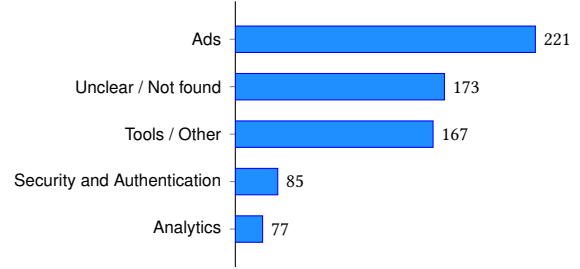

The automated exfiltration detection (\secref{sec:meth_automated_signal}) yields 723 SDKs that exhibit behavior similar to the known fingerprinting behavior of the SDKs in the Seed Set. Each SDK may have multiple versions and our SDK dataset identifies 14,178 versions for these 723 SDKs, for an average of 19.60 versions per SDK.
With the Extended Set labeled as described in \autoref{sec:meth_automated_signal} we consider the prevalence of various purposes across SDKs that exhibit fingerprinting-like behavior and plot the resulting distribution in \autoref{fig:purpose-stats}.

Several observations are readily available from the plot. First, there is a clear separation between the ``Ads'' category, the ``Tools / Other'' category, and the rest of the categories. The ``Tools / Other'' category being large is expected, as it encompasses a wide variety of functionalities, from cloud storage, to image and video processing, and to consent management. The high number of SDKs with ``Ads'' as purpose is potentially representative of the complex structure of that particular industry, where ad networks provide their own SDKs and ad platforms act as aggregators and mediators between apps and ad networks, with corresponding mobile SDKs that reflect these relationships. Oftentimes, the ad-mediation systems consists of 5--15 individual SDKs, one for each ad network to which the mediator can connect, easily boosting the number of total SDKs in this vertical. We opted to count these connector SDKs separately --- from a security and privacy perspective they are distinct artifacts.

A second observation is there is a large contingent of SDKs whose purpose is not clear from online, public information. The ``Unclear / Unfound'' category is the second highest source of likely fingerprinting behavior. We know these SDKs are in use in a variety of apps (as we see later in \secref{sec:rq3}), so the question is not only what purpose these SDKs serve, but also how to make this purpose information available to security and privacy enforcement mechanisms. One option would be to reverse engineer the code of each SDK and infer from this code its purpose, though this is unlikely to be a scalable long-term solution (and out of scope for this paper).

A further consideration for the use of privacy-preserving alternatives is the long tail of functionalities present in the ``Tools / Other'' category, which represents 23\% of the total number of likely fingerprinting SDKs. While ``Ads'', ``Security and Authentication'', and ``Analytics'' are reasonably well understood and studied in terms of privacy, the ``Tools / Other'' SDKs cover a broad range of algorithms and data types that may not have readily available privacy-preserving alternatives.

\paragraph{SDK Purpose \& Behavior.}
The challenge of identifying the purpose of likely fingerprinting SDKs of a particular type could be alleviated by focusing on their use of fingerprinting signals, if sufficiently discriminative. For example, if Ads SDKs have distinct fingerprinting behaviors (in terms of signals and API data they exfiltrate) compared to those of Security and Authentication SDKs, one could identify and control fingerprinting appropriately without relying on the SDK's purpose declaration or on its non-fingerprinting functionality, both of which could be adversarially manipulated. We evaluated this hypothesis by expressing the fingerprinting behavior of each SDK as points in a high-dimensional space defined by a one-hot encoding of the APIs exfiltrated. Each API of interest is an independent dimension in this space and an SDK is placed at position $0$ along this dimension if it does not exfiltrate the corresponding API, or $1$ if it does. This results in $504$-dimensional space (one for each API observed in the Extended Set of SDKs) in which we locate the $723$ SDKs.

\begin{figure*}
  \centering

  \Description{This is a scatter plot showing the distribution of data points in two dimensions. Each point represents an item and is plotted based on its coordinates. The points are colored and shaped differently to represent five distinct categories. A legend in the top right corner explains which shape and color correspond to which category. Here's a breakdown of the visual elements. Chart Type: Scatter plot. Data Points: Numerous data points are scattered across the plot area. There are no visible axes or grid lines, suggesting the plot represents a transformed space (like a t-SNE visualization). Categories and Markers: The data points are differentiated by their color and shape, according to the legend: Unclear / Unfound: Blue circles (o), Security & Authentication: Orange 'x' markers (x), Analytics: Green squares (s), Tools / Other: Red plus signs (+), Ads: Purple diamonds (d). Distribution: The points are not uniformly distributed. They appear to form several clusters or groupings in different areas of the plot. Some clusters contain points of multiple categories, while others are predominantly composed of points from a single category. Legend: A legend is located in the top right corner. It lists each category name alongside the color and shape of the marker used for that category in the plot. Overall, the scatter plot provides a visual representation of how data points from different categories are related or grouped in a two-dimensional space. The clustering of points of the same color and shape suggests that items within those categories share some similarities.}

  \includegraphics[width=\textwidth]{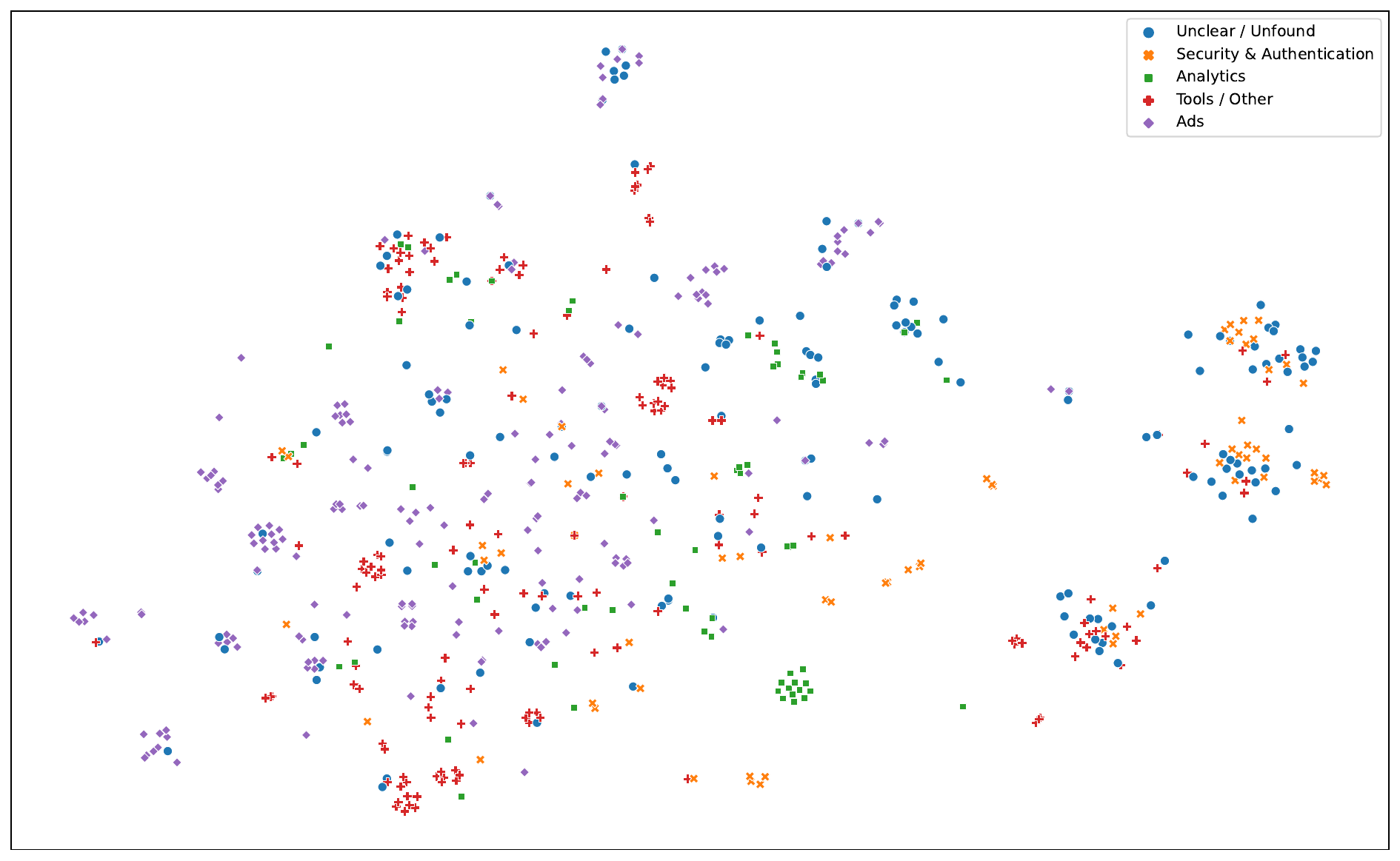}

  \caption{Map of likely fingerprinting behaviors of SDKs in the Extended Set, computed using t-SNE over embeddings constructed by one-hot encoding the exfiltrated APIs. Proximity of SDKs indicates that they exfiltrate data from similar sets of APIs. }
  \label{fig:t_sne}
\end{figure*}

In \autoref{fig:t_sne},
 we provide a visual representation of the similarity between different SDK types in the Extended Set using a t-Distributed Stochastic Neighbor Embedding (t-SNE)~\cite{JMLR:v9:vandermaaten08a} plot.
t-SNE allows us to cluster the SDKs based on their ``natural'' similarity in fingerprinting behavior by mapping the high-dimensional space (our 504 signals) to a faithful representation in a lower-dimensional space through non-linear transformations while preserving local and global relationships between the data points. 
Based on the recommendations from Wattenberg, Viégas, and Johnson~\cite{wattenberg2016how}, we set t-SNE perplexity to $25$, learning rate of $10$, and iterations to 5,000, resulting in a final KL divergence of $0.614789$. The resulting t-SNE output is shown in \autoref{fig:t_sne}, with one dot per SDK, relatively positioned as determined by t-SNE and color coded based on our five SDK purpose labels.

The t-SNE plot illustrates the diversity of signal/API usage in fingerprinting behavior, as a large number of small clusters formed. At a minimum this leads us to believe that a corresponding larger number of small, focused permission-based policies may be able to address the fingerprinting problem, with the associate risk of enforcement performance (due to the cost of maintaining and evaluating this many policies) and low usability (due to placing the user in the position of making decision based on seemingly similar but privacy-distinct permissions).

The feasibility of automating the anti-fingerprinting/anti-tracking policies put forth by the industry (advertising: no tracking allowed, anti-fraud: tracking allowed) can be reduced to whether ``Ads'' SDKs (marked as \textcolor{DarkViolet}{$\medblackdiamond$} in \autoref{fig:t_sne}) are easily separable from ``Security and Authentication'' SDKs (\textcolor{DarkOrange}{$\medblacktimes$} in \autoref{fig:t_sne}). The right third of t-SNE plot contains \textit{most} of the ``Security and Authentication'' SDKs (\textcolor{DarkOrange}{$\medblacktimes$}), while the ``Advertising'' SDKs (\textcolor{DarkViolet}{$\medblackdiamond$}) are on the left. Yet there are many ``Security and Authentication'' SDKs on the left side of the plot, not to mention ``Analytics'' (\textcolor{DarkGreen}{$\medblacksquare$}) and ``Tools / Other'' SDKs (\textcolor{Red}{$\medblackplus$}), that appear to have similar fingerprinting-like behavior to the ``Ads'' SDKs. Thus any automatic enforcement that needs to distinguish between ``Ads'' SDKs and ``Security and Authentication'' SDKs will need to rely on non-trivial classifiers that are more expressive than permissions.

Finally we observe that the ``Unclear / Unfound'' SDKs (\textcolor{DodgerBlue}{$\medblackcircle$}), which are declared in APKs and present in Maven repositories but lack any descriptive information, have fingerprinting-like behaviors similar to \textit{all} other SDK categories. This supports the need for robust categorization and labeling mechanisms for SDKs, and the conclusion that behavioral analysis may be insufficient without additional out-of-band (non-code) information.

\paragraph{Use of Sensitive Signals.}

We manually identified 24 APIs which could be used to retrieve location data exactly or approximately and then checked how many of the likely fingerprinting behaviors in SDKs from the Extended Set rely in these APIs. We performed a similar analysis for app-usage signals (based on the three APIs we identified to provide information about the apps the user has installed on the device, the apps that are in use, or the usage statistics for installed apps) and for the account-list signals (based on two APIs to retrieve lists of personal accounts registered on the device). 
We find that of likely fingerprinting SDKs $72\%$ collect coarse-grained location signals, $71.6\%$ collect fine-grained location, and $86.29\%$ collect at least one or the other. Only only $6.15\%$ record account-list signals, and $38.46\%$ collect app usage information.

\subsection{RQ3: Market Reach}
\label{sec:rq3}
\begin{quote}
    \textit{What kinds of apps use SDKs with fingerprinting behavior, and how prevalent are these SDKs in real-world apps?}
\end{quote}
To answer this question, we consider the SDK categories described in the RQ2 results, and the app categories assigned by the Google Play Store~\cite{google-play-categories}.

We are interested in understanding the presence of fingerprinting SDKs in the mobile-app marketplace. For this, we measured how many apps include fingerprinting SDKs in each app category, which categories of fingerprinting SDK are in most use, and which fingerprinting SDKs co-occur most often in apps. The first measurement seeks to determine whether there are app categories with particularly high prevalence of fingerprinting and thus that should be prioritized for any fingerprinting-reduction intervention. The second and third measurements inform any technical efforts to replace fingerprinting-based solutions with privacy-preserving alternatives.

\begin{figure*}
  \centering

  \Description{This image contains two horizontal bar charts placed side-by-side, along with a table of percentages on the right. Both bar charts share the same vertical axis, which lists various app categories. Left Bar Chart: Title/Axis Label: "Number of apps (in thousands)" is labeled along the horizontal axis, with tick marks indicating values from 0 to 350 in increments of 50. Bars: Each app category has a horizontal bar extending to the right from a central vertical line. Blue Bars: Represent the number of apps in each category. These bars are generally longer than the red bars, indicating a higher number of apps. Red Bars: These are very short bars next to the blue bars for each category. Their meaning isn't explicitly labeled, but they likely represent a related metric with a much smaller scale. Categories: The categories are listed vertically on the left side, from top to bottom: Comics, Libraries and Demo, Parenting, Weather, Dating, Video Players, Events, Art and Design, Beauty, House and Home, Photography, News and Magazines, Auto and Vehicles, Maps and Navigation, Medical, Social, Sports, Communication, Travel and Local, Personalization, Books and Reference, Finance, Music and Audio, Lifestyle, Shopping, Entertainment, Health and Fitness, Productivity, Food and Drink, Tools, Business, Game, Education. Trend: The blue bars generally get longer as you move down the list of categories, indicating that categories lower down have a higher number of apps. Right Bar Chart: Title/Axis Label: "Number of installs (in billions)" is labeled along the horizontal axis, with tick marks indicating values from 0 to 40 in increments of 20. The axis extends to the right. Bars: Similar to the left chart, each category has a horizontal bar. Blue Bars: Represent the number of installs in billions. These bars vary in length. Red Bars: Again, very short bars next to the blue bars. Categories: The categories are the same as the left chart and are implicitly linked by their vertical position. Trend: There is no obvious ascending or descending trend in the length of the blue bars in this chart; they vary significantly across categories. Table of Percentages: Location: To the right of the right bar chart. Content: Two columns of percentages are presented for each category. The percentages are highlighted with a yellow background. The first column of percentages appears to correspond to the blue bars in the left chart (number of apps). The second column of percentages appears to correspond to the blue bars in the right chart (number of installs). In essence, this image presents a comparison of app categories based on the number of apps and the number of installs, using two bar charts and a table of corresponding percentages. The left chart focuses on the quantity of apps per category, while the right chart focuses on the popularity or usage of apps per category (based on installs). The percentage table provides a relative measure for both metrics.}

  \subcaptionbox{Count of apps (blue) and apps with fingerprinting-like SDKs (red) by app category.\label{subf:app-counts}}[3.75in]{
  \begin{tikzpicture}
      \begin{axis}[
            height=5in,
            width=3.25in,
            xbar=0pt, 
            bar width=4pt,
            scaled x ticks=base 10:-3,
            ytick=data,
            axis y line=center,
            y axis line style={-},
            ymin = -0.5,
            ymax = 32.5,
            yticklabels from table={\allAppsTable}{AppCategory},
            y tick label style={anchor=east,font=\scriptsize\sffamily}, 
            axis x line=left,
            x tick label style={font=\scriptsize,color=black}, 
            xmin=0,
            enlarge x limits=upper,
            xlabel={Number of apps (in thousands)},
            xlabel style={font=\scriptsize\sffamily},
            x tick label style={font=\scriptsize\sffamily},
            xtick scale label code/.code={},
         ]
         \addplot+
            table [y expr=\coordindex,x=TotalAppsInCategory] {\allAppsTable};
         \addplot+
            table [y expr=\coordindex,x=AppsWithFingerprinters] {\allAppsTable};
      \end{axis}
  \end{tikzpicture}
  }
  \quad
  \subcaptionbox{Installs of apps (blue) and apps with fingerprinting SDKs (red), by app category.\label{subf:install-counts}}[1.50in]{
  \begin{tikzpicture}
      \begin{axis}[
            height=5in,
            width=1.50in,
            xbar=0pt, 
            bar width=4pt,
            scaled x ticks=base 10:-9,
            axis y line=none,
            ymin = -0.5,
            ymax = 32.5,
            yticklabels from table={\allAppsTable}{AppCategory},
            y tick label style={anchor=east,font=\scriptsize\sffamily}, 
            axis x line=left,
            x tick label style={font=\scriptsize\sffamily,color=black},
            xmin=0,
            enlarge x limits=upper,
            xlabel={Number of installs (in billions)},
            xlabel style={font=\scriptsize},
            xtick scale label code/.code={},
         ]
         \addplot+
            table [y expr=\coordindex,x=TotalInstallsInCategory] {\allAppsTable};
        \addplot+
            table [y expr=\coordindex,x=AppsWithFingerprintersInstalls] {\allAppsTable};
      \end{axis}
  \end{tikzpicture}
  }
  \quad
  \subcaptionbox{Percent of apps containing fingerprinting-like SDKs by total count (left) and install volume (right). \label{subf:prevalence}}[1.25in]{
     \newcommand{\OLDarraystretch}{\arraystretch}
     \renewcommand{\arraystretch}{1.35}
     \scriptsize\sffamily
     \begin{tabular}[b]{@{}rr@{}}
        \rowcolor{yellow}
		7.50\%	& 87.76\% \\
		1.58\%	& 5.58\% \\
		4.98\%	& 38.26\% \\
		5.71\%	& 15.78\% \\
        \rowcolor{yellow}
		4.09\%	& 82.57\% \\
		10.45\%	& 23.91\% \\
        \rowcolor{yellow}
		0.81\%	& 42.98\% \\
        \rowcolor{yellow}
		2.74\%	& 33.00\% \\
        \rowcolor{yellow}
		1.11\%	& 52.45\% \\
        \rowcolor{yellow}
		2.16\%	& 56.51\% \\
		5.69\%	& 22.27\% \\
        \rowcolor{yellow}
		3.85\%	& 45.40\% \\
		2.37\%	& 6.14\% \\
        \rowcolor{yellow}
		3.56\%	& 51.07\% \\
        \rowcolor{yellow}
		2.33\%	& 31.66\% \\
        \rowcolor{yellow}
		2.75\%	& 59.25\% \\
        \rowcolor{yellow}
		2.90\%	& 56.08\% \\
		2.17\%	& 10.63\% \\
		2.57\%	& 18.28\% \\
        \rowcolor{yellow}
		2.02\%	& 31.08\% \\
        \rowcolor{yellow}
		2.65\%	& 31.68\% \\
		4.30\%	& 37.48\% \\
        \rowcolor{yellow}
		2.11\%	& 32.80\% \\
        \rowcolor{yellow}
		2.22\%	& 33.74\% \\
        \rowcolor{yellow}
		2.71\%	& 54.44\% \\
        \rowcolor{yellow}
		3.46\%	& 42.21\% \\
        \rowcolor{yellow}
		1.77\%	& 31.27\% \\
        \rowcolor{yellow}
		1.72\%	& 26.71\% \\
        \rowcolor{yellow}
		1.10\%	& 63.80\% \\
		2.98\%	& 13.12\% \\
        \rowcolor{yellow}
		1.05\%	& 29.18\% \\
        \rowcolor{yellow}
		7.21\%	& 85.15\% \\
        \rowcolor{yellow}
		1.21\%	& 48.77\% \\
        \\
        \\
        \\[-0.25em]
      \end{tabular}%
      \renewcommand{\arraystretch}{\OLDarraystretch}%
  }
  \caption{The number of apps that come with fingerprinting SDKs is rather small, on average at less than 5\% of the total number of apps in a category (\subref{subf:app-counts}), yet these apps are some of the most installed (\subref{subf:install-counts}), giving likely fingerprinting SDKs an outsized presence in the market. In 23 app categories (\colorbox{yellow}{highlighted} in (\subref{subf:prevalence})), apps with fingerprinting SDKs are 10x more popular than other apps.}
  \label{fig:app-sdk-stats}
\end{figure*}
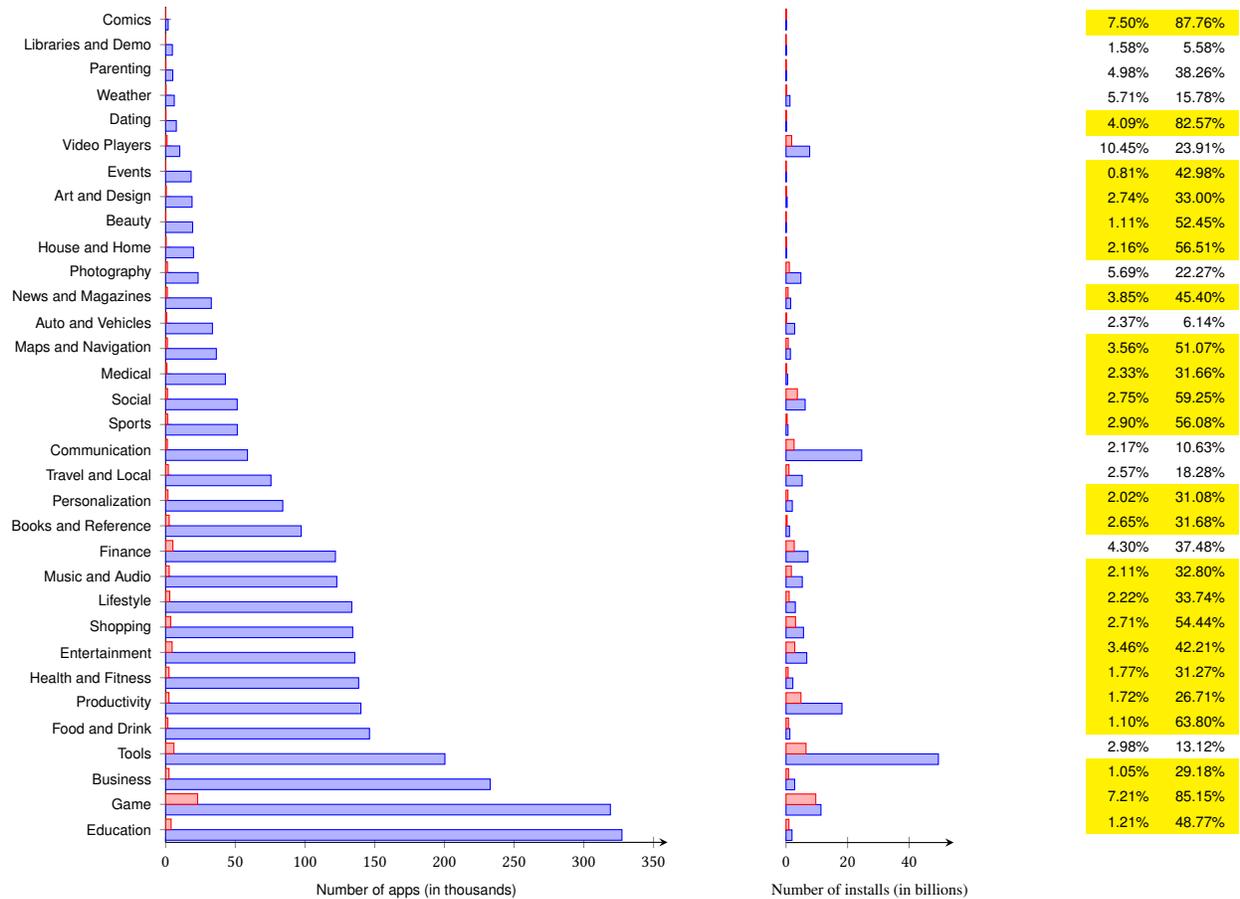

\autoref{fig:app-sdk-stats} shows the prevalence of apps that include fingerprinting SDKs in each app category. 
\autoref{subf:app-counts} illustrates that the raw number of apps using fingerprinting SDKs averages at 3.2\% across app categories, ranging between 0.8\% (for ``Events'' apps) and 10\% (for ``Video Players'' apps). \autoref{subf:install-counts} takes into account the number of installs each such app had in the 30-day period, and shows that presence of fingerprinting functionality is heavily skewed towards popular apps.

On average 39.4\% of apps in a category contain at least one fingerprinting SDK, and while the minimum prevalence is 5.5\% of apps (for the ``Libraries and Demo'' category), a number of app categories have $>$50\% prevalence: ``Maps and Navigation'', ``Beauty'', ``Shopping'', ``Sports'', ``House and Home'', ``Social'', ``Food and Drink'', ``Dating'', ``Game'', and ``Comics''. From a user's point of view, this indicates that randomly installing a popular app has a 39.4\% chance of being fingerprinted and, if they select a dating, game, or comics app, they will be fingerprinted with a 80+\% probability.

The ``Comics,'' ``Game,'' and  ``Dating'' categories stand out with the highest number of apps incorporating likely fingerprinting SDKs, respectively in this order. This could be attributed to several factors, such as the prevalence of free-to-use functionality (e.g., free-to-play games) that rely on targeted advertising or in-app purchases, or the need for (frictionless) user identification in online settings.

\pgfplotstableread[col sep=comma, trim cells, header=true]{data/app_categories_and_installs.data}\loadedAllAppsTable
\pgfplotstablesort[sort key={TotalAppsInCategory},sort cmp={int >}]\allAppsTable\loadedAllAppsTable

\begin{figure*}
    \centering

    \Description{This image contains two heatmaps side-by-side, along with a color bar to the left and a color bar to the right. Both heatmaps display relationships between categories, using color intensity to represent values. Left Heatmap: Shape: This heatmap is a vertical rectangle, taller than it is wide. Axes Labels: The vertical axis is labeled with several SDK categories rotated at an angle, including "Ads", "Analytics", "Security and Authentication", "Tools / Other", and "Unclear / Unfound". The horizontal axis is also labeled with the same categories in the same order. Color Scale: A color bar to the left of this heatmap shows a color gradient from light yellow at the bottom (representing lower values, around 0.0) to dark red at the top (representing higher values, up to 1.0). The tick marks on this color bar are at 0.2, 0.4, 0.6, 0.8, and 1.0. Cell Colors: The cells in this heatmap are colored according to the yellow-to-red color scale. The colors suggest varying levels of something being measured between the categories on the axes. There are distinct color variations across the grid. Right Heatmap: Shape: This heatmap is a triangle, specifically the upper right triangle of a square matrix. The lower left portion is blank or not displayed. Axes Labels: The vertical axis is labeled with a long list of app categories, including "Art and Design", "Auto and Vehicles", "Beauty", and many others, extending down the side. The horizontal axis is labeled with the same app categories, rotated at an angle along the bottom. Color Scale: A color bar to the right of this heatmap shows a color gradient from very light blue at the bottom (representing lower values, around 0.2) to dark blue at the top (representing higher values, up to 0.6). The tick marks on this color bar are at 0.2, 0.3, 0.4, 0.5, and 0.6. Cell Colors: The cells within the triangular area are colored according to the light blue-to-dark blue color scale. The intensity of the blue in each cell indicates a value related to the two corresponding app categories from the axes. There are noticeable patterns of darker blue (higher values) in certain areas of the triangle. Overall Interpretation: The image displays two distinct sets of relationships visualized as heatmaps. The left heatmap seems to show relationships between a smaller set of broad categories, using a yellow-to-red color scheme. The right heatmap, displaying only the upper triangle of a matrix, shows relationships between a much larger set of app categories, using a light blue-to-dark blue color scheme. The color bars provide the key to understanding the intensity of these relationships. The absence of the lower triangle in the right heatmap suggests that the relationship being measured is symmetrical (the relationship between Category A and Category B is the same as between Category B and Category A), and displaying both halves would be redundant.}
    
    \includegraphics[trim={0 .40cm 0 0}, clip, width=\textwidth]{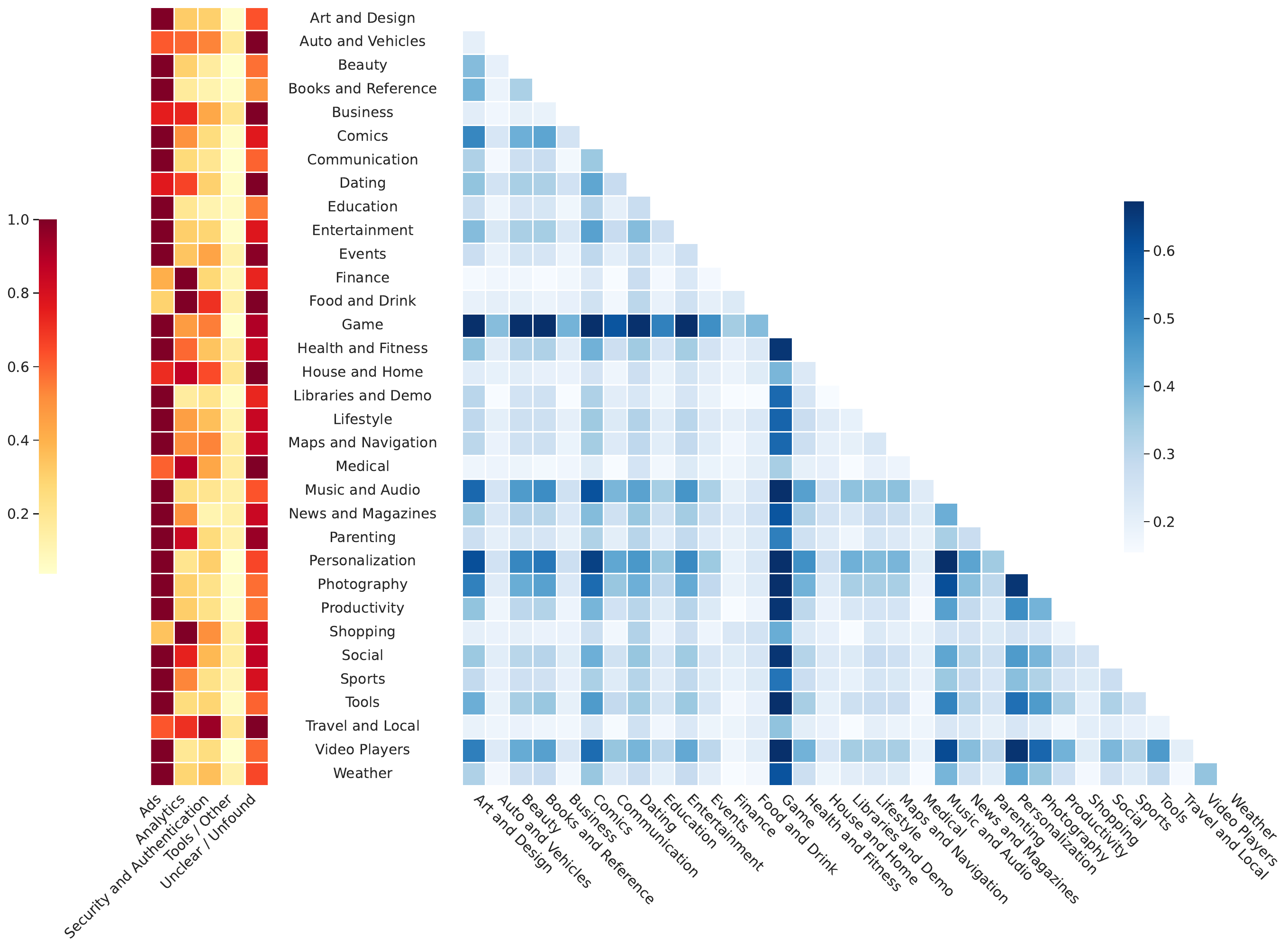}
    \subcaptionbox{Heatmap of the prevalence of likely fingerprinting SDKs in app categories (X-axis) and SDK categories (Y-axis). \label{subf:app_cats_sdk_cats}}[0.29\textwidth]{}
    \hspace*{0.01\textwidth}
    \subcaptionbox{Heatmap of the co-occurrence of likely fingerprinting SDKs across pairs of app categories. Each cell describes the percentage of apps in the corresponding app categories (on the X-axis and Y-axis) that share one or more likely fingerprinting SDKs. 
    \label{subf:app_app_probs}}[0.69\textwidth]{}
    \caption{Prevalence of likely fingerprinting SDKs within and across app categories.
             In~(\subref{subf:app_cats_sdk_cats}), the color gradient is computed per app category, allowing the comparison of SDK prevalence between app categories.
             In~(\subref{subf:app_app_probs}), the color gradient is computed per pair of app categories given by the X-axis and Y-axis coordinates. For raw data, see \autoref{sec:raw-RQ3}.
            }
    \label{fig:app_sdk_probs}
\end{figure*}

\paragraph{Fingerprinting Prevalence across App Categories}
To further understand the prevalence of likely-fingerprinting SDKs in the mobile-app ecosystem, we use the purpose labels we developed in \secref{sec:rq2} to map app categories to SDK categories. This results in the heatmap shown in \autoref{subf:app_cats_sdk_cats}, in which a deeper shade of red indicates that the SDK category of that row dominates the app category of that column. For example, any likely-fingerprinting SDKs used by ``Art and Design'' apps come primarily from the ``Ads'' SDK category, while likely-fingerprinting SDKs in ``Finance'' apps are foremost from the ``Analytics'' SDK category.

The ``Ads'' SDKs dominate across almost all app categories as a source of likely fingerprinting behavior, with ``Unclear / Unfound'' SDKs as the second most common. We note that in many cases the absolute number of ``Unclear / Unfound'' SDKs is close to that of ``Ads'' SDKs (e.g., in the ``Business'' app category, the labels for 16,097 SDKs are unclear, and 17,780 SDKs have the ``Ads'' label) and thus any shift from ``Unclear / Unfound'' to ``Ads'' will only further cement the dominance of ``Ads'' SDKs a source of likely fingerprinting behavior.

A second observation from this heatmap is that there are several app categories (``Finance'', ``Food and Drink'', ``Shopping'') where ``Analytics'' likely-fingerprinting SDKs are more prevalent that other categories of likely-fingerprinting SDKs. We hypothesize that in these app categories, the fingerprinting is less used to track user identities (which are known from the user-account information) and more used to understand user preferences with respect to the items for sale in the app.

\paragraph{Possible Tracking across App Categories via Fingerprint Sharing}
A significant privacy risk brought on by fingerprinting is the potential for a third party to track user activity across applications. 
This can happen when an SDK included in multiple applications fingerprints a user, and thus allows for user activity to be attributed to the same user for both. 
A service might then learn, say, that a user that engages in particular style of dating app, and also uses a specific medical or finance application. 
Such cross-app tracking may take place on device or on the server, in both cases powered by the fingerprinting data obtained from the shared SDK.

To estimate a \textit{lower bound} on the risk of cross-app tracking, we analyze the prevalence of likely fingerprinting SDKs present in distinct categories by computing the probability that two apps randomly selected from each app category share at least one likely fingerprinting SDK. The results are shown in \autoref{subf:app_app_probs} as a heatmap (only the lower diagonal presented, as the heatmap is symmetric). A darker shade of blue in the figure indicates a higher prevalence of shared fingerprinting SDKs, as shown, for example, by the (``Game'', ``Entertainment'') entry compared with the (``Travel and Local'', ``Comics'') entry. We compute these probabilities for the top-1000 apps by total audience size (as defined in \secref{sec:meth_app_data}) for each app category, regardless of whether those apps include a likely fingerprinting SDK or not. As a result, the prevalence (and the associated heatmap shown in \autoref{subf:app_app_probs}) reflect both the popularity of apps and the distribution of likely fingerprinting SDKs in such popular apps.

Analysis of this heatmap clearly indicates that a few app categories have likely fingerprinting SDKs in common with many other app categories. For example, ``Game'' apps share likely fingerprinting SDKs with ``Art and Design'' apps, as well as with apps in the ``Beauty'', ``Books and Reference'', ``Comics'', ``Communication'', ``Dating'', ``Entertainment'', ``Health and Fitness'', ``Libraries and Demo'', ``Lifestyle'', ``Maps and Navigation'', ``Music and Audio'', ``News and Magazines'', ``Personalization'', ``Photography'', ``Productivity'', ``Social'', ``Tools'', ``Video Players'', and ``Weather'' categories. Similarly, ``Personalization'' apps share SDKs with 7 out of 33 app categories. From a user's point of view, this implies that they are at higher risk of cross-app tracking if they install apps from both the ``Game'' and ``Comics'' categories.

Alternatively, some categories of apps rarely share likely fingerprinting SDKs with other app categories. We highlight the ``Finance'' and ``Medical'' app categories in particular, as such apps often process highly sensitive data. Without further study, one cannot tell why fingerprinting is not more common here, and we note that such apps often require the user to authenticate to access their bank or investment account or their medical record and as such may not need to fingerprint the user through indirect signals.

\section{Discussion \& Limitations}
\label{sec:discussion}

Our results indicate that the fingerprinting ecosystem is more complex than previously estimated, both in terms of fingerprinting behavior and fingerprinting purpose. We interpret the results below, discuss limitations of our methodology, and present implications for app security mechanisms.

\paragraph{Challenges for Sector-Specific Solutions.} A core observation of this paper is that the current mobile ecosystem has evolved to organically deploy fingerprinting-like behavior in a wide variety of apps through several types of SDKs. Our analysis reveals that while advertising SDKs contribute to fingerprinting, they are not the sole culprits: A significant portion of fingerprinting-like behavior originates from SDKs employed for analytics and anti-fraud purposes, and a large contingent ($23.9\%$) did not have sufficient public information about their purpose or functionality to discern a category. These SDKs are often integrated for reasons directly outside of monetization --- understanding user behavior to improve applications or preventing bots and fraud --- despite ultimately collecting enough device information to create a fingerprint. 

This finding challenges the prevailing notion that fingerprinting is primarily driven by app developer's need to monetize via advertising, and highlights the opportunity for a broader perspective on privacy-preserving alternatives. Research on how developers can be incentivized to adopt privacy preserving analytics, for example, could prove useful.  
It is also likely that sandboxing efforts (such as Android's Privacy Sandbox \cite{PrivacySandboxTechnologya}) could provide additional benefit for both detection and enforcement against SDKs that over-collect --- though systems research into lightweight sandboxing to support this setting is necessary, as scaling current process-based techniques to non-ads SDKs requires untenable overhead for constrained mobile environments.

\paragraph{Challenges for API-Specific or Other Behavioral Defenses.} 
A surprising result of our exploration of the Seed Set (self-identified fingerprinting SDKs) was that the space of APIs used is sparse; the SDKs collected information from dissimilar sets of APIs. This held true regardless of the use-case of the SDK or of its prevalence in our dataset of applications. 
A potential explanation might be API Proxying~\cite{jha2023formalanalysisapiproxy}, which may further complicate any anti-fingerprinting enforcement, though determining the joint entropy or shared entropy of particular API's is beyond the scope of this work.

In any case, it would appear that targeting specific APIs (akin to Apple's required reasons~\cite{AppleRequiredReason}) represents a brittle defense against fingerprinting. 
Developers have a number of signals at their disposal, and could easily move to other sources of entropy. Further, though we (surprisingly) found no evidence of non-API hardware-based fingerprinting in our Seed Set, one might expect developers to shift more advanced methods if comprehensive enforcement at the API level were introduced.

\paragraph{Potential for Sector-Specific Analysis \& Targeting.} 
It is worth noting that certain sensitive application verticals appeared to have an improved privacy stance. 
Normalized by install volume, only 30\% of applications in the medical category used a fingerprinting SDK, and (assuming a normal distribution holds between sample sets) only 19.5\% of \textit{those} did so using an ads SDK --- the bulk of identifiable fingerprinting behavior appears to come from analytics. Medical applications also appeared to have a lower potential for cross-application tracking. 
This heartening result, which is largely repeated in the Finance category, highlights the need for future work focusing on solutions for specific sensitive market verticals. 

\paragraph{Need for Multi-Platform Analysis.} 
It is likely that our results extend to the iOS ecosystem --- indeed, all SDKs in our Seed Set appear to have versions readily available for iOS --- a finding consistent with prior work on cross-platform tracking~\cite{kollnigAreIPhonesReally2022}. 
However, it is difficult to perform such analysis on iOS, as Apple's application and operating-system wide DRM restricts third parties' ability to scalably perform static and dynamic analysis on applications in their App Store. Future work studying the iOS ecosystem would provide invaluable insight into the effectiveness of design choices between the two operating systems.

\subsection{Limitations}

Any empirical study, including the present paper, is a limited view into real-world conditions and trends and thus it is important to evaluate the factors that threaten its validity. Following the ``Campbell Tradition''~\cite{campbell2015experimental}, we consider four types of validity---internal, statistical, construct, and external---and their impact on this study.

\emph{Internal validity} refers to whether the measured effect (fingerprintable APIs) truly corresponds to the outcome of interest (fingerprinting behavior). A risk is that the use of APIs to retrieve high-entropy data may not be caused by intentional fingerprinting behavior, but instead the result of necessary app functionality. 
We sidestep this by focusing on documenting the purposes of collection of fingerprintable data. A secondary limitation exists in the selection bias implicit in our Seed Set, which consists of SDKs that \textit{self-identify} as fingerprinting. 
It may be that SDKs that fingerprint for hidden reasons use alternative techniques, which would not be caught in our later analyses. 
We assume that a Seed Set SDK's self-reporting is honest and make no further inferences about the SDK's intent.

\emph{Statistical validity} refers to the risks of underpowered experiments, i.e., without sufficient statistical support. Our large sample size of 228,598 SDKs and 3,025,417 apps mitigates this risk.

\emph{Construct validity} refers to the choice of metrics to measure the presence of fingerprintable APIs and behaviors. We focused on the number of APIs as an efficient metric of fingerprinting behavior, though we note that not all APIs are equally useful for fingerprinting. For now we make the simplifying assumption that in-the-wild techniques are largely equivalent, and that there is no relationship between signals collected. Using more complex metrics such as collision entropy~\cite{bravohermsdorff2023privatecommunicationefficientalgorithmsentropy} requires experiments across large sets of devices and users, which we leave for future work.

\emph{External validity} refers to the generalizability of our results to real-world. Our choice of actual SDKs from popular Maven repositories and mobile apps from the Google Play store ensure minimize this risk. However, we do not attempt to catalog all fingerprinting mobile ecosystem, limiting ourselves to Java-language SDKs that are part of the Android/AOSP framework (excluding non-platform APIs or those from OEMs). 
It is possible that the Seed Set of fingerprinting SDKs, hand selected through web search, is not representative of all fingerprinting behaviors in the wild, and further study to ensure a comprehensive view is needed.

\section{Conclusion}

In this paper, we presented the largest-scale analysis of SDK behavior ever conducted, examining over 228,000 SDKs and 178,000 Android applications to understand the prevalence and purpose of fingerprinting-like behavior. Our findings reveal that a significant number of SDKs, beyond those explicitly designed for advertising, collect enough information to potentially track users. This includes SDKs used for analytics and anti-fraud, highlighting the need for privacy-preserving alternatives in these areas.  Surprisingly, a large portion of SDKs exhibiting fingerprinting-like behavior lacked clear identification, emphasizing the need for greater transparency in the SDK ecosystem.  Moreover, we observed that these SDKs with fingerprinting-like behavior are disproportionately popular and often integrated across diverse application categories. These results underscore the importance of ongoing efforts by Apple and Google to enhance user privacy and emphasize the need for continued research to ensure that such industry efforts are well directed.

\bibliographystyle{plain}
\bibliography{paper,sdk_ident}

\appendix
{ \sffamily
\begin{tcolorbox}[colback=green!5!white,colframe=green!50!black,title=Conference Version]
A shorter version of this paper is published at the \href{https://www.sigsac.org/ccs/CCS2025/}{ACM Conference on Computer and Communications Security 2025 (CCS'25)}. The present version additionally includes:
\begin{itemize}[leftmargin=*]
    \item Descriptions of the static analyses performed,
    \item Description of the SDK-identification algorithm,
    \item Definition of the codebook used to categorized SDKs, and
    \item List of the APIs observed in fingerprinting-like behaviors.
\end{itemize}
\end{tcolorbox}
}

\section{Data Tables for Fingerprinting Prevalence in SDK and App Categories}
\label{sec:raw-RQ3}

The following tables provide detailed data on our market measurements.

\autoref{tab:fp-prev-sdk+app-cat} presents the prevalence of likely fingerprinting SDKs across various app categories, broken down by SDK type. For instance, in the ``Art and Design'' app category, 43.3\% of apps are likely to contain Ads SDKs that likely engage in fingerprinting. The data shows that ``Ads'' and ``Unclear/Unfound'' SDK categories generally have higher prevalence rates across most app categories compared to ``Analytics,'' ``Security and Authentication,'' and ``Tools/Other'' SDKs.

Tables~\ref{tab:fp-co-occurr-prev-app-cat-1-of-2} and~\ref{tab:fp-co-occurr-prev-app-cat-2-of-2} detail the proportion of apps within each category that contain SDKs also present in apps of other categories. The tables indicate that many apps utilize SDKs that are also prevalent in apps belonging to different categories. For instance, while 0.401 of ``Books and Reference'' apps share SDKs with ``Art and Design'' apps, only 0.095 of ``Art and Design'' apps share SDKs with the ``Food and Drink'' category, indicating a much lower overlap in SDK usage between these two specific app types.

\begin{table}[ht]
    \caption{The prevalence of likely fingerprinting SDKs (by SDK category) in app categories.}
    \label{tab:fp-prev-sdk+app-cat}

    \centering
    \pgfkeys{/pgf/number format/.cd,fixed,fixed zerofill,precision=3}
    \small
    \begin{filecontents*}{data/FP-SDK-prevalence-in-app-categories-percentages.csv}
Category,Ads,Analytics,Security and Authentication,Tools / Other,Unclear / Unfound
Art and Design,0.4332939787,0.1381345927,0.132231405,0.02243211334,0.2739079103
Auto and Vehicles,0.2114537445,0.2004405286,0.1828193833,0.06387665198,0.3414096916
Beauty,0.4835680751,0.1455399061,0.07981220657,0.01408450704,0.2769953052
Books and Reference,0.5442199404,0.09307717787,0.06624710169,0.02981119576,0.2666445843
Business,0.2412280702,0.2335526316,0.1381578947,0.06688596491,0.3201754386
Comics,0.3867924528,0.1933962264,0.09905660377,0.02358490566,0.2971698113
Communication,0.4747826087,0.1269565217,0.09565217391,0.01826086957,0.2843478261
Dating,0.2753164557,0.2373417722,0.1075949367,0.02215189873,0.3575949367
Education,0.5162775403,0.1016113121,0.06149292996,0.03584347254,0.2847747451
Entertainment,0.4118060436,0.1277231202,0.1189388616,0.02125790583,0.3202740689
Events,0.3455882353,0.1176470588,0.1544117647,0.04411764706,0.3382352941
Finance,0.1625,0.396969697,0.1109848485,0.03787878788,0.2916666667
Food and Drink,0.09464082098,0.3181299886,0.2246294185,0.04446978335,0.3181299886
Game,0.3388280133,0.1599428299,0.1849642687,0.009509290138,0.3067555979
Health and Fitness,0.3402417962,0.2003454231,0.1174438687,0.05526770294,0.286701209
House and Home,0.2086614173,0.2519685039,0.188976378,0.05905511811,0.2913385827
Libraries and Demo,0.4625,0.075,0.1,0.025,0.3375
Lifestyle,0.35818409,0.1649312786,0.1291128696,0.04414827155,0.3036234902
Maps and Navigation,0.3251282051,0.1671794872,0.1743589744,0.05128205128,0.2820512821
Medical,0.1952380952,0.2880952381,0.1404761905,0.05238095238,0.3238095238
Music and Audio,0.4519708029,0.1080291971,0.09313868613,0.06306569343,0.2837956204
News and Magazines,0.3861386139,0.1930693069,0.04455445545,0.05198019802,0.3242574257
Parenting,0.3144329897,0.2628865979,0.0824742268,0.0412371134,0.2989690722
Personalization,0.4511612022,0.09357923497,0.1403688525,0.01673497268,0.2981557377
Photography,0.4642289348,0.1399046105,0.1033386328,0.02437731849,0.2681505034
Productivity,0.4630095622,0.1454453951,0.105183694,0.02767991948,0.2586814293
Shopping,0.1210295728,0.3472070099,0.1774370208,0.05585980285,0.2984665936
Social,0.3172268908,0.2359943978,0.1197478992,0.05112044818,0.2759103641
Sports,0.3752455796,0.199410609,0.08546168959,0.03831041257,0.3015717092
Tools,0.4535585042,0.1164723227,0.1297413215,0.02908457311,0.2711432784
Travel and Local,0.179462572,0.2034548944,0.2715930902,0.05854126679,0.2869481766
Video Players,0.4828496042,0.09234828496,0.1207124011,0.01781002639,0.2862796834
Weather,0.4113924051,0.1181434599,0.1476793249,0.05274261603,0.2700421941
\end{filecontents*}
\pgfplotstabletypeset[col sep=comma,
                          header=has colnames,
                          every nth row={3}{before row=\arrayrulecolor{lightgray}\midrule\arrayrulecolor{black}},
                          every head row/.style={before row={\toprule & \multicolumn{5}{c}{SDK Category} \\ \cmidrule{2-6}},
                                                 after row=\midrule,
                                                 typeset cell/.code={
            \ifnum\pgfplotstablecol=\pgfplotstablecols
            \pgfkeyssetvalue{/pgfplots/table/@cell content}{\rotatebox{65}{##1}\\}%
            \else
            \ifnum\pgfplotstablecol=1
            \pgfkeyssetvalue{/pgfplots/table/@cell content}{##1&}%
            \else
            \pgfkeyssetvalue{/pgfplots/table/@cell content}{\rotatebox{65}{##1}&}%
            \fi
            \fi
                                                     }
                              },
                          every last row/.style={after row=\bottomrule},
                          columns/Category/.style={string type,column type=l,column name=App Category},
                          columns/{Ads,Analytics,Tools / Other,Unclear / Unfound}/.style={numeric type},
                          columns/Security and Authentication/.style={numeric type,column name=Sec. and Authn},
                          ]
       {data/FP-SDK-prevalence-in-app-categories-percentages.csv}
\end{table}

\setlength\rotFPtop{0pt plus 1fil}
\begin{sidewaystable*}
    \caption{The sharing prevalence of likely fingerprinting SDKs across app categories [Part 1]. Part~2 is in \autoref{tab:fp-co-occurr-prev-app-cat-2-of-2}.}
    \label{tab:fp-co-occurr-prev-app-cat-1-of-2}
    \pgfkeys{/pgf/number format/.cd,fixed,fixed zerofill,precision=3}
    \centering
    \small
    \begin{filecontents*}{data/FP-SDK-co-occurrence-prevalence-in-app-categories.csv}
category,Art and Design,Auto and Vehicles,Beauty,Books and Reference,Business,Comics,Communication,Dating,Education,Entertainment,Events,Finance,Food and Drink,Game,Health and Fitness,House and Home,Libraries and Demo,Lifestyle,Maps and Navigation,Medical,Music and Audio,News and Magazines,Parenting,Personalization,Photography,Productivity,Shopping,Social,Sports,Tools,Travel and Local,Video Players,Weather
Art and Design,0.4691233523,0.2023514892,0.3809206963,0.4007753843,0.2098861432,0.498664301,0.3221982043,0.3646954622,0.2721309939,0.3806892376,0.2718851979,0.1636189338,0.195014746,0.7941528015,0.3657139542,0.2135706857,0.3040550883,0.296044323,0.3011313108,0.1770868453,0.5608598541,0.3411906682,0.2691111835,0.6102507225,0.5114875075,0.363934809,0.202363174,0.3510825508,0.2897395051,0.4123510048,0.1919918907,0.5146799241,0.3216198666
Auto and Vehicles,0.2023514892,0.1819390771,0.1974930952,0.1865517745,0.1744159369,0.2384230626,0.1662922858,0.2534551893,0.178208229,0.232939824,0.193256708,0.1739217423,0.203116804,0.377899226,0.211329505,0.1965864985,0.1434235083,0.2034404363,0.1895196431,0.1819955918,0.2469212184,0.2315490071,0.203073801,0.2541850056,0.2181798644,0.1776244587,0.1904395538,0.2126913775,0.1988715108,0.1926540182,0.1771150586,0.2219782023,0.1691577265
Beauty,0.3809206963,0.1974930952,0.3321130145,0.3274043886,0.1999737682,0.409808319,0.2688461452,0.3312610526,0.2425104648,0.3304132001,0.2494106925,0.1735149675,0.2033204789,0.676762907,0.3130991725,0.2114379221,0.2533404973,0.2648682517,0.2611431893,0.1831790462,0.4572664371,0.3086310029,0.2482797288,0.498958431,0.4171759388,0.29906218,0.2025431781,0.3052967632,0.263772281,0.3340906888,0.1904016944,0.4210560295,0.2678624876
Books and Reference,0.4007753843,0.1865517745,0.3274043886,0.3465150741,0.1924477869,0.435411455,0.2779584048,0.3264920752,0.2409770056,0.3352782316,0.241962264,0.1582393529,0.1863421727,0.7140959705,0.3216184683,0.1984771982,0.2602065494,0.2646581668,0.2652694749,0.1681952402,0.4878288901,0.3062190396,0.2424031841,0.5295818071,0.4422963293,0.3149029642,0.1904335665,0.3099875309,0.2590433252,0.3562835153,0.178997024,0.4448863261,0.2789381247
Business,0.2098861432,0.1744159369,0.1999737682,0.1924477869,0.1713194238,0.250548552,0.1685623139,0.2540959177,0.176848519,0.2361735522,0.1865581401,0.1727141276,0.1977674348,0.4026696639,0.2137781517,0.18946325,0.1446243478,0.202963175,0.1880917353,0.1742755291,0.2601201948,0.230500712,0.1997378335,0.268962264,0.2310912052,0.1828155106,0.1902020589,0.2166675476,0.1980211454,0.2004705038,0.1722620783,0.2343759586,0.1721634043
Comics,0.498664301,0.2384230626,0.409808319,0.435411455,0.250548552,0.5769699248,0.3517207723,0.4319771527,0.3087862483,0.442933498,0.2959885642,0.2256731467,0.2564299651,0.8472723027,0.4068155017,0.250618238,0.3217866386,0.3485417462,0.33775191,0.2168311788,0.6052990975,0.3814880144,0.3191920127,0.6397392295,0.5533625071,0.3981996759,0.27337125,0.4108218079,0.3284918353,0.4569515543,0.2365750104,0.5528068641,0.3553450448
Communication,0.3221982043,0.1662922858,0.2688461452,0.2779584048,0.1685623139,0.3517207723,0.2283097835,0.2784361248,0.2026492093,0.2786153514,0.2064899222,0.1454530314,0.1715539073,0.5976171021,0.2663730502,0.1776394993,0.2098926977,0.2237790555,0.2228050728,0.15343514,0.3947968226,0.2602157794,0.2079361112,0.4304242141,0.3563341152,0.2552231856,0.171819844,0.2570827303,0.2197400993,0.2872063403,0.1602826424,0.358224856,0.2266113473
Dating,0.3646954622,0.2534551893,0.3312610526,0.3264920752,0.2540959177,0.4319771527,0.2784361248,0.4188670423,0.2766752842,0.3810949691,0.2740169503,0.2743827294,0.3022076665,0.6686316455,0.3444259298,0.2666769626,0.2371028394,0.3182349613,0.295257305,0.2469362488,0.4400335861,0.3550730976,0.3076753936,0.4646280706,0.4098332116,0.3064100209,0.317525384,0.3572922555,0.3123741958,0.3417092301,0.261134763,0.3978276116,0.2757726622
Education,0.2721309939,0.178208229,0.2425104648,0.2409770056,0.176848519,0.3087862483,0.2026492093,0.2766752842,0.198845548,0.2673102656,0.2054828752,0.1665009928,0.1935560103,0.5123525987,0.2473971389,0.1922302173,0.1834541526,0.2213666909,0.211743931,0.1722613703,0.3347530682,0.2567189126,0.2141193877,0.355125771,0.2994565319,0.2240520293,0.1891435248,0.2451145886,0.2188881228,0.2483632135,0.174361703,0.3037344454,0.2042710894
Entertainment,0.3806892376,0.232939824,0.3304132001,0.3352782316,0.2361735522,0.442933498,0.2786153514,0.3810949691,0.2673102656,0.3810795845,0.2633042773,0.2311259234,0.2613334715,0.7135576562,0.3370559639,0.2494689415,0.2436605952,0.3034687033,0.2865678251,0.2248088193,0.4720158642,0.3392623662,0.2899833095,0.4943036019,0.4225273088,0.3103723404,0.2666500124,0.3448859922,0.2941156595,0.3523213984,0.2314017393,0.4272519073,0.2832449221
Events,0.2718851979,0.193256708,0.2494106925,0.241962264,0.1865581401,0.2959885642,0.2064899222,0.2740169503,0.2054828752,0.2633042773,0.2274265804,0.1654104049,0.2012332552,0.4825635891,0.251297878,0.2097118585,0.192310592,0.2267370581,0.2196681903,0.188453653,0.3276469756,0.2652478562,0.2216604465,0.3501648681,0.2906408532,0.2251218368,0.1820754114,0.241704625,0.2247814361,0.2445736736,0.1840396208,0.2990673996,0.2094184681
Finance,0.1636189338,0.1739217423,0.1735149675,0.1582393529,0.1727141276,0.2256731467,0.1454530314,0.2743827294,0.1665009928,0.2311259234,0.1654104049,0.2101884158,0.225178784,0.3376994064,0.19742202,0.185287376,0.1084902297,0.2016304809,0.1744578726,0.1783204954,0.2010402874,0.2215018819,0.2004294476,0.1970933746,0.1914340135,0.155253761,0.2353418112,0.21731878,0.196811537,0.1696307465,0.1849511455,0.1779592156,0.1442590052
Food and Drink,0.195014746,0.203116804,0.2033204789,0.1863421727,0.1977674348,0.2564299651,0.1715539073,0.3022076665,0.1935560103,0.2613334715,0.2012332552,0.225178784,0.2533259463,0.3808114465,0.2270164099,0.2169263678,0.1336632971,0.2299489666,0.2053607474,0.2077394462,0.2388074014,0.2542857236,0.2292292767,0.2365608419,0.2204708383,0.1820255642,0.2534733453,0.2426166987,0.2238205068,0.197619344,0.2108163599,0.2121165015,0.1704855779
Game,0.7941528015,0.377899226,0.676762907,0.7140959705,0.4026696639,0.8472723027,0.5976171021,0.6686316455,0.5123525987,0.7135576562,0.4825635891,0.3376994064,0.3808114465,0.9936380673,0.6586185094,0.3939910997,0.5572344331,0.5678540573,0.560237945,0.3338767346,0.8860312322,0.5989151393,0.5164362406,0.9123793819,0.8377177196,0.6609544173,0.4162332446,0.6611003241,0.5350266907,0.7384059279,0.3659832914,0.8405354546,0.603376706
Health and Fitness,0.3657139542,0.211329505,0.3130991725,0.3216184683,0.2137781517,0.4068155017,0.2663730502,0.3444259298,0.2473971389,0.3370559639,0.251297878,0.19742202,0.2270164099,0.6586185094,0.3201128461,0.2269073357,0.2427086688,0.2758975828,0.2683114267,0.2002052647,0.4452744272,0.3182136548,0.2604952734,0.4787413694,0.4037261074,0.2978763576,0.2285945591,0.3117193861,0.2717407693,0.333045319,0.2078321345,0.4053350588,0.2683318075
House and Home,0.2135706857,0.1965864985,0.2114379221,0.1984771982,0.18946325,0.250618238,0.1776394993,0.2666769626,0.1922302173,0.2494689415,0.2097118585,0.185287376,0.2169263678,0.3939910997,0.2269073357,0.2166196455,0.1540836137,0.2185825589,0.2029460061,0.1982505319,0.2633642019,0.2505604181,0.2185006875,0.2682503006,0.2282137989,0.1896213999,0.1989684582,0.2262565342,0.2143944568,0.2047632791,0.1903552317,0.2380652421,0.1833917621
Libraries and Demo,0.3040550883,0.1434235083,0.2533404973,0.2602065494,0.1446243478,0.3217866386,0.2098926977,0.2371028394,0.1834541526,0.2436605952,0.192310592,0.1084902297,0.1336632971,0.5572344331,0.2427086688,0.1540836137,0.2116635393,0.1959365193,0.1995668032,0.1303055826,0.3689653987,0.2365601464,0.1809472022,0.4077923062,0.3324883276,0.2365109421,0.1276711779,0.2237311163,0.1951939094,0.264142583,0.1335049973,0.3370200618,0.2097976226
Lifestyle,0.296044323,0.2034404363,0.2648682517,0.2646581668,0.202963175,0.3485417462,0.2237790555,0.3182349613,0.2213666909,0.3034687033,0.2267370581,0.2016304809,0.2299489666,0.5678540573,0.2758975828,0.2185825589,0.1959365193,0.2538710281,0.237613967,0.1991032199,0.3664885875,0.2849370582,0.2439319343,0.3841251436,0.3290801746,0.2472028697,0.2284993193,0.2797773712,0.245559009,0.2763645702,0.2028063358,0.3298369645,0.2265990922
Maps and Navigation,0.3011313108,0.1895196431,0.2611431893,0.2652694749,0.1880917353,0.33775191,0.2228050728,0.295257305,0.211743931,0.2865678251,0.2196681903,0.1744578726,0.2053607474,0.560237945,0.2683114267,0.2029460061,0.1995668032,0.237613967,0.2322001113,0.181954527,0.3691489016,0.2727828669,0.2268913497,0.3964212104,0.3309752565,0.246970306,0.201399939,0.2632706753,0.2327479475,0.2751020692,0.1858025443,0.33312645,0.2240885657
Medical,0.1770868453,0.1819955918,0.1831790462,0.1681952402,0.1742755291,0.2168311788,0.15343514,0.2469362488,0.1722613703,0.2248088193,0.188453653,0.1783204954,0.2077394462,0.3338767346,0.2002052647,0.1982505319,0.1303055826,0.1991032199,0.181954527,0.1870052067,0.2181896039,0.2246766613,0.2002769335,0.2171330422,0.1908342099,0.1621928251,0.192695043,0.2042793928,0.1941110223,0.1741414481,0.1788409975,0.1959109308,0.1579184222
Music and Audio,0.5608598541,0.2469212184,0.4572664371,0.4878288901,0.2601201948,0.6052990975,0.3947968226,0.4400335861,0.3347530682,0.4720158642,0.3276469756,0.2010402874,0.2388074014,0.8860312322,0.4452744272,0.2633642019,0.3689653987,0.3664885875,0.3691489016,0.2181896039,0.669481807,0.4133523528,0.3317299614,0.7113529062,0.6112039189,0.4458225083,0.247265475,0.4336706143,0.3528525083,0.5052246399,0.2338082091,0.6183306933,0.3984363232
News and Magazines,0.3411906682,0.2315490071,0.3086310029,0.3062190396,0.230500712,0.3814880144,0.2602157794,0.3550730976,0.2567189126,0.3392623662,0.2652478562,0.2215018819,0.2542857236,0.5989151393,0.3182136548,0.2505604181,0.2365601464,0.2849370582,0.2727828669,0.2246766613,0.4133523528,0.3391907393,0.2759503132,0.4377010029,0.3732691017,0.2872517037,0.2511440414,0.3115422784,0.2879171277,0.3132630576,0.2291911019,0.377869337,0.2616193797
Parenting,0.2691111835,0.203073801,0.2482797288,0.2424031841,0.1997378335,0.3191920127,0.2079361112,0.3076753936,0.2141193877,0.2899833095,0.2216604465,0.2004294476,0.2292292767,0.5164362406,0.2604952734,0.2185006875,0.1809472022,0.2439319343,0.2268913497,0.2002769335,0.3317299614,0.2759503132,0.2405018321,0.3437911494,0.2967268228,0.2275985072,0.2245842116,0.2658611186,0.2385068283,0.2523340963,0.2012126236,0.299384058,0.2115476656
Personalization,0.6102507225,0.2541850056,0.498958431,0.5295818071,0.268962264,0.6397392295,0.4304242141,0.4646280706,0.355125771,0.4943036019,0.3501648681,0.1970933746,0.2365608419,0.9123793819,0.4787413694,0.2682503006,0.4077923062,0.3841251436,0.3964212104,0.2171330422,0.7113529062,0.4377010029,0.3437911494,0.7649326395,0.6592054959,0.4847697703,0.2497619466,0.4572520923,0.3733555075,0.5452931373,0.2388602056,0.6630437319,0.4312810184
Photography,0.5114875075,0.2181798644,0.4171759388,0.4422963293,0.2310912052,0.5533625071,0.3563341152,0.4098332116,0.2994565319,0.4225273088,0.2906408532,0.1914340135,0.2204708383,0.8377177196,0.4037261074,0.2282137989,0.3324883276,0.3290801746,0.3309752565,0.1908342099,0.6112039189,0.3732691017,0.2967268228,0.6592054959,0.5669744461,0.4033227413,0.2385881004,0.3955259094,0.3196991584,0.4577989664,0.2121451777,0.5644366712,0.3545824266
Productivity,0.363934809,0.1776244587,0.29906218,0.3149029642,0.1828155106,0.3981996759,0.2552231856,0.3064100209,0.2240520293,0.3103723404,0.2251218368,0.155253761,0.1820255642,0.6609544173,0.2978763576,0.1896213999,0.2365109421,0.2472028697,0.246970306,0.1621928251,0.4458225083,0.2872517037,0.2275985072,0.4847697703,0.4033227413,0.2890492662,0.1856704551,0.2867512631,0.2424380425,0.3258514443,0.172283627,0.4051747182,0.2557521691
Shopping,0.202363174,0.1904395538,0.2025431781,0.1904335665,0.1902020589,0.27337125,0.171819844,0.317525384,0.1891435248,0.2666500124,0.1820754114,0.2353418112,0.2534733453,0.4162332446,0.2285945591,0.1989684582,0.1276711779,0.2284993193,0.201399939,0.192695043,0.247265475,0.2511440414,0.2245842116,0.2497619466,0.2385881004,0.1856704551,0.2747088811,0.2515299672,0.2237558368,0.2053814188,0.2072883783,0.2171795469,0.1656709772
Social,0.3510825508,0.2126913775,0.3052967632,0.3099875309,0.2166675476,0.4108218079,0.2570827303,0.3572922555,0.2451145886,0.3448859922,0.241704625,0.21731878,0.2426166987,0.6611003241,0.3117193861,0.2262565342,0.2237311163,0.2797773712,0.2632706753,0.2042793928,0.4336706143,0.3115422784,0.2658611186,0.4572520923,0.3955259094,0.2867512631,0.2515299672,0.3220009634,0.2712503075,0.3244979316,0.2142496396,0.3932225086,0.2585973017
Sports,0.2897395051,0.1988715108,0.263772281,0.2590433252,0.1980211454,0.3284918353,0.2197400993,0.3123741958,0.2188881228,0.2941156595,0.2247814361,0.196811537,0.2238205068,0.5350266907,0.2717407693,0.2143944568,0.1951939094,0.245559009,0.2327479475,0.1941110223,0.3528525083,0.2879171277,0.2385068283,0.3733555075,0.3196991584,0.2424380425,0.2237558368,0.2712503075,0.2483915848,0.2656511977,0.1990952311,0.3211283362,0.2203311836
Tools,0.4123510048,0.1926540182,0.3340906888,0.3562835153,0.2004705038,0.4569515543,0.2872063403,0.3417092301,0.2483632135,0.3523213984,0.2445736736,0.1696307465,0.197619344,0.7384059279,0.333045319,0.2047632791,0.264142583,0.2763645702,0.2751020692,0.1741414481,0.5052246399,0.3132630576,0.2523340963,0.5452931373,0.4577989664,0.3258514443,0.2053814188,0.3244979316,0.2656511977,0.3723988529,0.186822517,0.4596683607,0.289606514
Travel and Local,0.1919918907,0.1771150586,0.1904016944,0.178997024,0.1722620783,0.2365750104,0.1602826424,0.261134763,0.174361703,0.2314017393,0.1840396208,0.1849511455,0.2108163599,0.3659832914,0.2078321345,0.1903552317,0.1335049973,0.2028063358,0.1858025443,0.1788409975,0.2338082091,0.2291911019,0.2012126236,0.2388602056,0.2121451777,0.172283627,0.2072883783,0.2142496396,0.1990952311,0.186822517,0.1807907522,0.2090223361,0.1608182753
Video Players,0.5146799241,0.2219782023,0.4210560295,0.4448863261,0.2343759586,0.5528068641,0.358224856,0.3978276116,0.3037344454,0.4272519073,0.2990673996,0.1779592156,0.2121165015,0.8405354546,0.4053350588,0.2380652421,0.3370200618,0.3298369645,0.33312645,0.1959109308,0.6183306933,0.377869337,0.299384058,0.6630437319,0.5644366712,0.4051747182,0.2171795469,0.3932225086,0.3211283362,0.4596683607,0.2090223361,0.5739299913,0.3621090314
Weather,0.3216198666,0.1691577265,0.2678624876,0.2789381247,0.1721634043,0.3553450448,0.2266113473,0.2757726622,0.2042710894,0.2832449221,0.2094184681,0.1442590052,0.1704855779,0.603376706,0.2683318075,0.1833917621,0.2097976226,0.2265990922,0.2240885657,0.1579184222,0.3984363232,0.2616193797,0.2115476656,0.4312810184,0.3545824266,0.2557521691,0.1656709772,0.2585973017,0.2203311836,0.289606514,0.1608182753,0.3621090314,0.2332376441
\end{filecontents*}
\pgfplotstabletypeset[col sep=comma,
                          columns={[index]0,[index]1,[index]2,[index]3,[index]4,[index]5,[index]6,[index]7,[index]8,[index]9,[index]10,[index]11,[index]12,[index]13,[index]14,[index]15,[index]16,[index]17},
                          header=has colnames,
                          every nth row={3}{before row=\arrayrulecolor{lightgray}\midrule\arrayrulecolor{black}},
                          every head row/.style={before row={\toprule & \multicolumn{17}{c}{App Category} \\ \cmidrule{2-18}},
                                                 after row=\midrule,
                                                 typeset cell/.code={
            \ifnum\pgfplotstablecol=\pgfplotstablecols
            \pgfkeyssetvalue{/pgfplots/table/@cell content}{\rotatebox{65}{##1}\\}%
            \else
            \ifnum\pgfplotstablecol=1
            \pgfkeyssetvalue{/pgfplots/table/@cell content}{##1&}%
            \else
            \pgfkeyssetvalue{/pgfplots/table/@cell content}{\rotatebox{65}{##1}&}%
            \fi
            \fi
                                                     }
                              },
                          every last row/.style={after row=\bottomrule},
                          columns/category/.style={string type,column type=l,column name={App Category}},
                          ]
       {data/FP-SDK-co-occurrence-prevalence-in-app-categories.csv}
\end{sidewaystable*}

\begin{sidewaystable*}
    \caption{The sharing prevalence of likely fingerprinting SDKs across app categories [Part 2]. Part~1 of this data is in \autoref{tab:fp-co-occurr-prev-app-cat-1-of-2}.}
    \label{tab:fp-co-occurr-prev-app-cat-2-of-2}

    \pgfkeys{/pgf/number format/.cd,fixed,fixed zerofill,precision=3}
    \centering
    \small
    \pgfplotstabletypeset[col sep=comma,
                          columns={[index]0,[index]18,[index]19,[index]20,[index]21,[index]22,[index]23,[index]24,[index]25,[index]26,[index]27,[index]28,[index]29,[index]30,[index]31,[index]32},
                          header=has colnames,
                          every nth row={3}{before row=\arrayrulecolor{lightgray}\midrule\arrayrulecolor{black}},
                          every head row/.style={before row={\toprule & \multicolumn{15}{c}{App Category} \\ \cmidrule{2-16}},
                                                 after row=\midrule,
                                                 typeset cell/.code={
            \ifnum\pgfplotstablecol=\pgfplotstablecols
            \pgfkeyssetvalue{/pgfplots/table/@cell content}{\rotatebox{65}{##1}\\}%
            \else
            \ifnum\pgfplotstablecol=1
            \pgfkeyssetvalue{/pgfplots/table/@cell content}{##1&}%
            \else
            \pgfkeyssetvalue{/pgfplots/table/@cell content}{\rotatebox{65}{##1}&}%
            \fi
            \fi
                                                     }
                              },
                          every last row/.style={after row=\bottomrule},
                          columns/category/.style={string type,column type=l,column name={App Category}},
                          ]
       {data/FP-SDK-co-occurrence-prevalence-in-app-categories.csv}
\end{sidewaystable*}

\section{Details of Our Static Analyses}
\label{sec:det-details}

\subsubsection*{Dependency Analysis for Android SDKs}

It is a common practice for a target software development kit (SDK) to rely on other SDKs. For instance, an advertising SDK may gather user data for fingerprinting and then share the data externally using another SDK (e.g., OkHttp). We refer to the target SDK as the \textit{main SDK} and the SDKs utilized by the main SDK as its \textit{dependency SDKs}. It is crucial to accurately infer and integrate the dependencies of each main SDK into the analysis for thorough detection.

To deduce the dependencies for every primary SDK, we analyze the Maven Project Object Model (POM) files of all the SDKs in our repository, extracting the referenced SDKs in each file. From each SDK's POM file information we construct a dependency graph with directed edges whenever one SDK's POM file references another's POM file. In the dependency graph, a distinct SDK version is represented by a trio of factors: a group ID indicating the developer, an artifact ID specifying the SDK name, and a version number, typically written as $X:Y:Z$ for version $Z$ of SDK $Y$ from developer $X$. During the resolution of the dependency graph, version conflict could happen. For instance, an SDK $M:A$ may reference both SDKs $N:B:1$ and $P:C:1$. However, SDK $N:B:1$ may require SDK $P:C:2$. This means that for SDK $P:C$, both versions $1$ and $2$ are listed as dependencies of SDK $M:A$. Since importing both versions into the static analysis could lead to one version overriding the other, causing non-deterministic behaviors, we only keep one version for each SDK. Based on Ockham's razor principle, we prioritize the version with the shortest path from the main SDK in the dependency graph, which is $P:C:1$ in the above example.

During the analysis, we configure the public methods in the main SDK as the analysis’ entry points, meaning that only the execution paths and behaviors triggered by one of these public methods will be reported.

\subsubsection*{Static Taint Analysis}

We employ static taint flow analysis to uncover potential instances of fingerprinting. This process begins by tainting PII data with metadata that allows the system to track its movement throughout the program's code.  To achieve this, the static taint flow analysis encompasses several key phases.  First, the system scans the input program to pinpoint likely taint sources (API method call where the program accesses PII) and sinks (API method calls where data exits the device).

Once sources and sinks are established, taint flags are propagated to construct a taint flow graph. This graph utilizes nodes to signify program elements (such as registers or fields), while edges encapsulate the possible transfer of tainted data. The graph expands incrementally until it reaches a fixed point.  A detected taint flow path connecting a source and sink flags potential PII exfiltration.

\begin{figure}
    \centering
    \Description{This image is a diagram illustrating data flow through app code using interconnected nodes and arrows. Here's a description of the elements. Nodes: There are several nodes of different shapes and colors, representing code locations and their role in the data flow. Green Rectangles with Rounded Corners: Two nodes labeled "Entry Point" are located at the top of the diagram. Purple Circles: Three circular nodes are present. A legend at the bottom of the diagram indicates that purple circles represent "Taint source". Light Green Circles: Seven circular nodes are scattered throughout the middle and bottom portions of the diagram. These nodes are not explicitly labeled in the legend but are part of the flow. Orange Circles: Two circular nodes are located at the bottom of the diagram. A legend at the bottom indicates that orange circles represent "Taint sink". Connections (Arrows): Dark gray arrows connect the nodes, indicating the direction of the flow. The arrows start from one node and point to another, showing the movement or progression through the diagram. Legend: A small legend is positioned at the bottom center of the diagram. It clearly defines the meaning of the colored circular nodes: Purple circle: "Taint source", Orange circle: "Taint sink". Flow Description: The arrows show complex paths connecting the nodes. The "Entry Point" nodes at the top lead to both purple ("Taint source") and light green nodes. The flow progresses through a network of light green and purple nodes, eventually reaching the orange ("Taint sink") nodes at the bottom. The arrows indicate various possible paths through the system. In summary, the diagram is a visual representation of an data flow through app code, highlighting specific points as "Entry Point," "Taint source," and "Taint sink," and showing the connections and possible paths between these points using arrows.}
    \includegraphics[width=0.55\columnwidth]{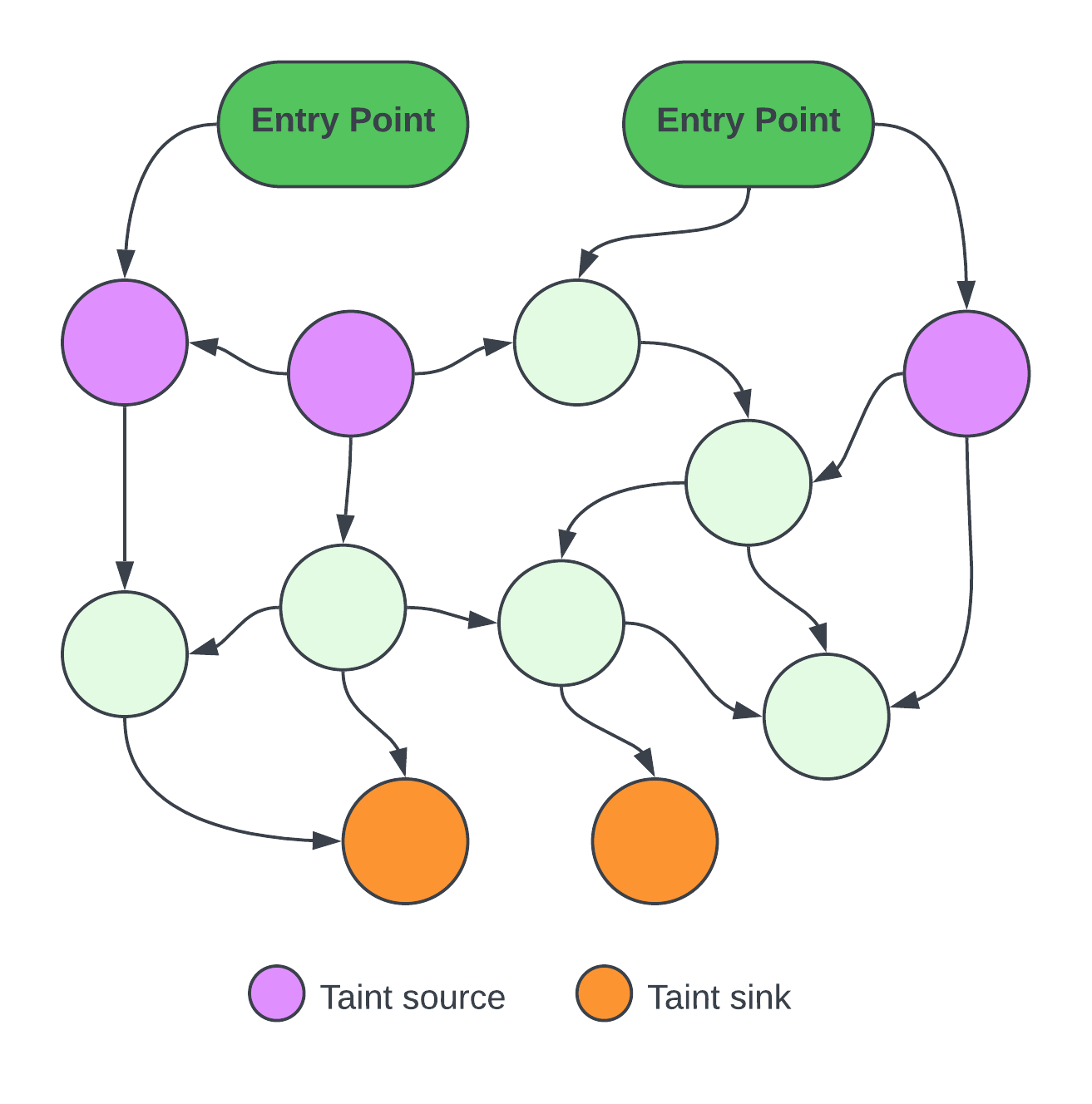}
    \caption{Static Taint Analysis -- Identify Sources and Sinks}
    \label{fig:taint_phase_1}
\end{figure}

\begin{figure}
    \centering
    \Description{This image is a diagram illustrating the propagation of taint labels through a data flow graph. Here's a description of the elements. Nodes: There are several nodes of different shapes and colors, representing code locations in the flow. Green Rectangles with Rounded Corners: Two nodes labeled "Entry Point" are located at the top of the diagram. Red Circles: Several circular nodes are colored red. A legend at the bottom indicates that red nodes and arrows represent "Taint flow". Light Green Circles: Five circular nodes are scattered throughout the middle and bottom portions of the diagram. These nodes are not explicitly labeled by color in the legend but are part of the overall structure. Connections (Arrows): Dark gray arrows connect some nodes, showing a general flow. However, some connections are highlighted in red, indicating a specific path or flow being tracked. These red arrows also represent the "Taint flow" according to the legend. Highlighted Flow (Red): A specific path through the diagram is highlighted in red. This path starts from the "Entry Point" on the left and proceeds through a sequence of red circular nodes connected by red arrows, ending at a red circular node near the bottom left. Another red path originates from the "Entry Point" on the right and follows red arrows through a sequence of red circular nodes, ending at a red circular node near the bottom right. Legend: A small legend is positioned at the bottom center of the diagram. It shows two red circles connected by a red arrow and is labeled "Taint flow". In summary, the diagram visually represents the propagation of taint labels through a data flow graph. It specifically highlights certain nodes and paths in red to illustrate a "Taint flow," making it easy to follow these particular routes through the system.}
    \includegraphics[width=0.55\columnwidth]{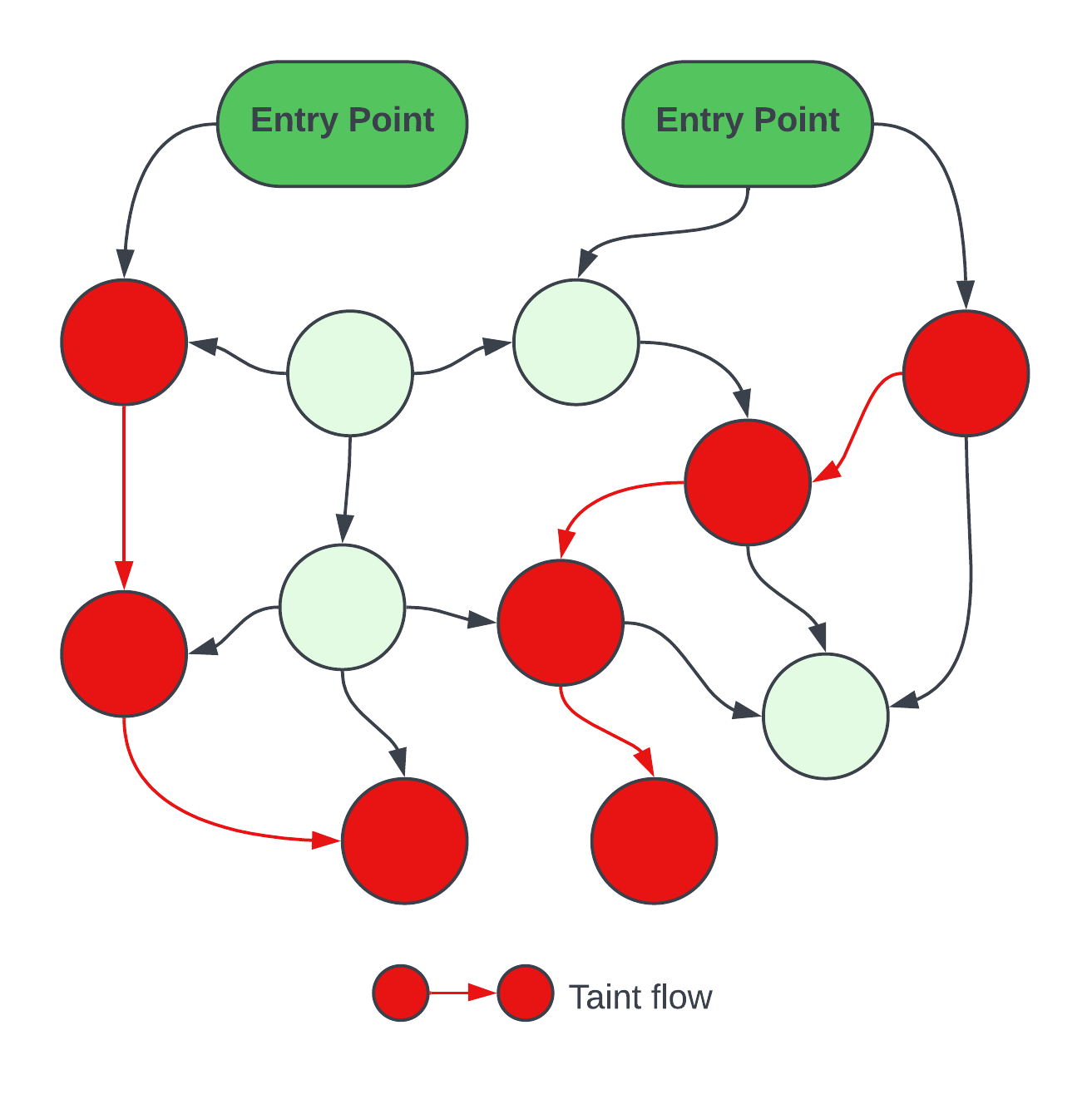}
    \caption{Static Taint Analysis -- Taint Propagation}
    \label{fig:taint_phase_2}
\end{figure}

\subsubsection*{CoFlow Analysis}

CoFlow analysis is a specialized taint flow analysis designed to identify suspicious app behavior patterns based on ``crossover flows''. Unlike the traditional taint flow analysis, which focuses on one-to-one relationships between data sources and sinks, CoFlow analysis detects scenarios where multiple sources converge into a single sink. %
This focus on many-to-one relationships makes it useful for uncovering behaviors like Fingerprinting or ID bridging.

CoFlow analysis builds upon the taint-flow tracking capabilities, utilizing our static taint analysis process to follow the movement of data throughout an application. It applies a set of constraints and rules to pinpoint those flow patterns that match the configured suspicious behavior definitions. CoFlow analysis uses a dedicated configuration file to define its behavior detection rules. This file specifies sets of source APIs and a corresponding sink API and includes fine-grained constraints to minimize false positives.

The analysis process centers on ensuring that, for each configured behavior, at least one source from each defined source group participates in that behavior. The system examines potential sinks, traces the flow of data backward, and compares discovered sources against the rule set. If a match is found for every source group within a rule, the system flags this as an instance of the suspicious behavior. Optimizations exist to streamline this process and improve efficiency. When CoFlow analysis detects a suspicious behavior, it generates an output including a set of sources (with one source representing each configured source group) and the associated sink.

\begin{figure}
    \centering
    \Description{This image is a diagram illustrating a coflow analysis process for app code. Here's a description of the elements. Nodes: There are several nodes of different shapes and colors, representing code locations. Green Ovals: Two nodes labeled "Entry point" are located at the top of the diagram. Blue Circles: Three circular nodes are present. A legend at the bottom of the diagram indicates that blue circles represent "Taint source". Light Green Circles: Seven circular nodes are scattered throughout the middle and bottom portions of the diagram. These nodes are not explicitly labeled in the legend but are part of the flow. Yellow Circle: One circular node is located towards the bottom center of the diagram. A legend at the bottom indicates that a yellow circle represents "Taint sink". Connections (Dashed Arrows): Dashed gray arrows connect the nodes, indicating the direction of the flow. The arrows start from one node and point to another, showing the movement or progression through the diagram. The dashed lines suggest a non-solid or potential connection. Legend: A small legend is positioned at the bottom center of the diagram. It clearly defines the meaning of the colored circular nodes: Blue circle: "Taint source", Yellow circle: "Taint sink". Flow Description: The dashed arrows show various paths connecting the nodes. The "Entry point" nodes at the top lead to both blue ("Taint source") and light green nodes. The flow progresses through a network of light green and blue nodes, eventually leading to the single yellow ("Taint sink") node. In summary, the diagram is a visual representation of a coflow analysis process, highlighting specific points as "Entry point," "Taint source," and "Taint sink." It uses dashed arrows to show potential or implied connections between these points.}
    \includegraphics[width=\columnwidth]{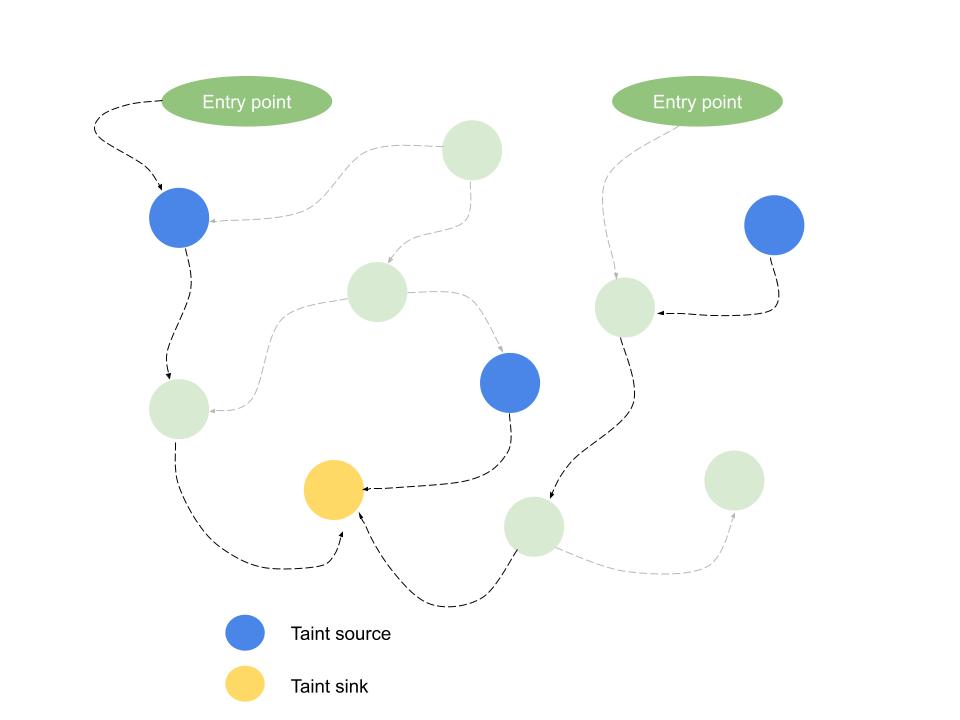}
    \caption{CoFlow Analysis}
    \label{fig:coflow1}
\end{figure}

\subsubsection*{Fingerprinting Detection}

Fingerprinting detection leverages CoFlow analysis to identify fingerprinting behaviors in SDKs. A ``fingerprinting behavior'' is defined as a set of $N$ or more fingerprinting sources that flow to a common interesting sink. The analysis proceeds from the list of fingerprinting APIs observed in our Seed Set of SDKs by checking that at least $N$ of the data items from fingerprinting APIs are exfiltrated after possibly being combined into new data objects. The CoFlow source group for fingerprinting is configured with the set of signals collected by self-reported fingerprinting SDKs. The number of sources is quite large, and we expect that most fingerprinting SDKs will include only a subset of these sources in their identifiers. Fingerprinting sinks are configured in two groups: network APIs, which create the potential for fingerprint exfiltration over the network, and encryption functions. Other categories of sinks exist but were not included in this work. For example, fingerprinting sources may be collected in a map or JSON object and simply returned by a publicly visible SDK method. Similarly, an SDK may populate an argument that is passed by reference to a public method with fingerprinting sources.

Fingerprinting diverges slightly from other CoFlow use cases in that we derive higher confidence in the behavior being present as more CoFlows are detected. Therefore, it's useful to include all detected sources in the output for fingerprinting behaviors, instead of the usual one source per group. Including all detected sources also allows for more thorough analysis of fingerprinting behaviors across the corpus of SDKs, as well as useful information for debugging and reverse engineering to confirm the behavior.

\subsubsection*{Challenges}
We run our analysis on SDKs bundled with their dependencies, and their dependencies' dependencies, and so on. Our first iteration of fingerprinting detection included taint sources from the entire SDK and dependency bundle. Because these bundles can become quite large, and taint tracking exhibits superlinear behavior with an increasing number of sources, we found that our analysis was prohibitively expensive for very large SDKs and SDKs with a high number of dependencies.

Additionally, tracking taint sources that originate in dependencies leads to identifying fingerprinting behaviors that are fully contained within a dependency, and potentially not utilized by the main SDK at all. This "over-detection" was very noisy, and we saw many SDKs being flagged for including the same popular fingerprinting SDKs as dependencies. Identifying which SDKs include fingerprinting SDKs as dependencies is valuable, but using expensive static analysis techniques to do it is not efficient, and the noise may obscure fingerprinting performed by the main SDK.

To solve for these issues, we reduced the scope of taint sources to only those originating in the main SDK, excluding all sources that originate in dependencies. Notably, taint \textit{sinks} are still included if they are in a dependency, allowing us to catch cases where an SDK collects fingerprinting sources but uses a dependency to hash or exfiltrate them. We acknowledge that this trade-off means we will miss some cases of fingerprinting, mainly where an SDK uses dependencies that each fetch a number of sources below the threshold for fingerprinting, but that when combined in the main SDK \textit{do} reach the threshold. The ability to detect boundaries of the SDK that are not interesting and can be excluded from analysis, as well as some dynamic analysis techniques, may be useful in filling this gap.

\begin{figure}
    \centering
    \Description{This image is a diagram illustrating a coflow analysis process of app code and SDK code. The diagram is spread out horizontally and features multiple distinct groupings of nodes. Here's a description of the elements. Nodes: There are several nodes of different shapes and colors, some with labels (numbers or letters). Green Ovals: Two nodes labeled "Entry point" are located towards the top, one on the left and one in the upper middle. Blue Circles with Numbers: There are four blue circular nodes with numbers inside them: node 1 (upper left), node 2 (middle left), node 3 (upper middle), and node 6 (lower right). Yellow Circles with Letters: There are two yellow circular nodes with letters inside them: node A (lower left) and node B (lower right). Light Green Circles (Unlabeled): Numerous unlabeled light green circular nodes are scattered throughout the diagram, serving as intermediate points in the flow. Connections (Dashed Arrows): Dashed gray arrows connect the nodes, indicating the direction of the flow. These dashed lines suggest potential or implied connections. Groupings and Flow: The diagram appears to show several separate or semi-separate flows. Left Group: Starts from the left "Entry point", connects to node 1 (blue), then through light green nodes to node 2 (blue), and potentially towards node A (yellow) and other light green nodes. Middle Group: Starts from the middle "Entry point", connects to node 3 (blue), then through light green nodes to nodes 4 (blue) and 5 (blue), and also branches out to other light green nodes. Right Group: A separate cluster of light green nodes in the lower right corner, with node 6 (blue) and node B (yellow) connected within this group. Legend: Although not explicitly formatted as a legend box, there are colored circles with text labels below the left group of nodes, serving as a key: Blue circle: Taint source, Yellow circle: Taint sink. In summary, this diagram presents a coflow analysis process for app code and SDK code. It categorizes certain nodes as "Entry point", "Taint source" (blue circles with numbers), and "Taint sink" (yellow circles with letters). The dashed arrows indicate potential paths, and the nodes are arranged in a way distinguish between app code and SDK code.}
    \includegraphics[width=\columnwidth]{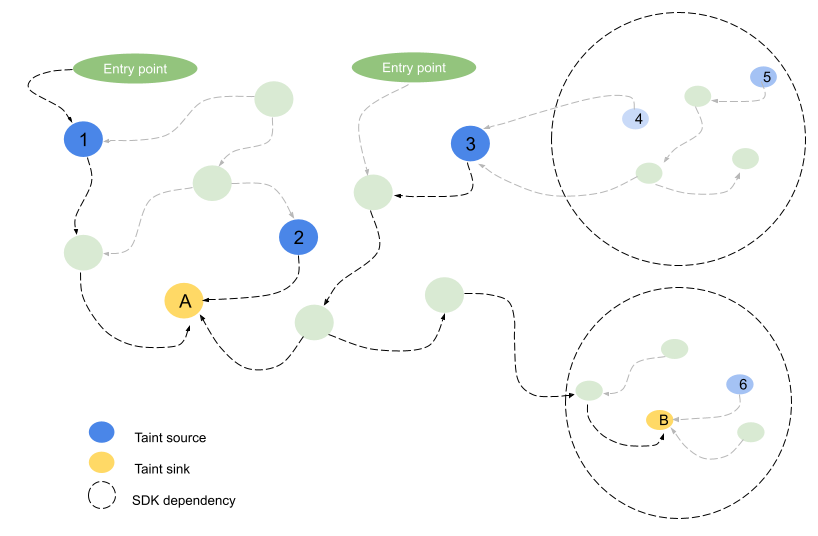}
    \caption{Fingerprinting detection using CoFlow. There are 2 fingerprinting behaviors in this example. Sources \{1,2,3\} \textrightarrow SinkA, and source \{3\} \textrightarrow SinkB. Sources 4, 5, and 6 are not included in the results because they are in a dependency.}
    \label{fig:coflow2}
\end{figure}

\section{Details of Our SDK Identification Approach}
\label{sec:sdk-ident-details}

Any approach for SDK identification must not rely on having access to source code (since apps are primarily distributed in binary format), must not assume all SDK code is present in the app (since compilation and linking often minify, optimize, or remove SDK code), must handle code obfuscation (prevalent in Android apps, with more than 75\% of them use obfuscation), must handle SDK dependencies (in 30,000 Java SDKs we analyzed, the 80 percentile of SDKs depended on 17 other SDKs), and must operate at scale (the number of SDKs and versions is on the order of $10^4$--$10^7$ as the data from \url{modulecounts.com} shows, with growth rate of $10^2$--$10^3$ new SDKs per day).

Our solution to SDK identification uses a fine-grained similarity metric for code that can be aggregated across code units (e.g., methods, classes, packages) and distribution units (e.g., SDKs, SDK families). %
The similarity metric for methods is based on a similarity across the following set of features: the type signature of the method, the Java- and Android-framework APIs invoked in the method body, the string constants used by the method body, and the histogram of instruction types present in the method body.
\begin{itemize}
    \item The type signature of the method captures the types of its parameters and the type of its return value. Since all types outside of the Java and Android frameworks are defined by the developer, their names are not trustworthy and cannot be relied upon for the purpose of similarity comparison. We eliminate all programmer-chosen type names to obtain an ``anonymized'' type signature for the method. The resulting type signature is one-hot encoded as a (boolean-valued) feature to be embedded into our vector space.
    \item The Java- and Android-framework APIs invoked in the method body each becomes one feature, counting the number of invocations present in the method body. Such a feature does not account for indirect ways to invoke the framework APIs, for example via Java reflection, via code in other languages (e.g., native or Javascript code), or via dynamic code loaded at runtime.
    \item The string constants used in the method body each form one (boolean-valued) feature, similar to the anonymized type signature of the method.
    \item The instruction types present in the method body each form one feature, measuring the frequency of that specific instruction type in the method body. We consider 34 types of instructions, closely matching the instructions in the Dalvik VM specification, ranging from \textsf{ASSIGN}, to \textsf{LOAD\_INSTANCE}, and to \textsf{THROW}.
\end{itemize}

The feature vector $\mathit{fv}(m) \in \mathbb{R}^{1\times d}$ for a method $m$ is derived by combining all of the above features and embedding them into a space of $2^{64}$ dimensions by hashing each feature name using a 64-bit hash function. The hash function needs to be collision resistant, to ensure that the adversary cannot easily create code similar to some target SDK code and evade SDK identification, but not necessarily pre-image resistant and thus does not need to necessarily be a cryptographic hash. This embedding step constructs a sparse vector representation for each Java method, since Java methods typically have only hundreds of non-zero features.

\begin{figure}
   \centering
   \Description{This is a vertical bar chart displaying the number of features based on their frequency. Here's a breakdown of the elements. Chart Type: Vertical bar chart. Horizontal Axis (X-axis): Labeled "Feature Frequency". This axis represents how many times a feature appears. The labels are integers from 1 to 10. Vertical Axis (Y-axis): Labeled "Number of Features". This axis represents the count of features. The scale ranges from 0 to 12,500,000, with tick marks at 2,500,000, 5,000,000, 7,500,000, 10,000,000, and 12,500,000. Bars: Blue vertical bars rise from the horizontal axis. Each bar corresponds to a "Feature Frequency" value on the X-axis, and its height indicates the "Number of Features" with that frequency on the Y-axis. Data Representation: The bar at "Feature Frequency" 1 is the tallest, reaching over 11,000,000. The bar at "Feature Frequency" 2 is significantly shorter, around 4,500,000. The bars at higher feature frequencies generally decrease in height, with a small increase at frequency 10. The bars for frequencies 7, 8, and 9 are very short, close to the horizontal axis. Grid Lines: Faint gray horizontal grid lines extend from the Y-axis tick marks across the chart area, aiding in reading the height of the bars. In summary, the bar chart visually shows the distribution of feature frequencies. It highlights that a very large number of features appear only once, and the number of features decreases dramatically as their frequency increases, with a slight uptick in features appearing 10 times.}
   \includegraphics[width=0.75\columnwidth]{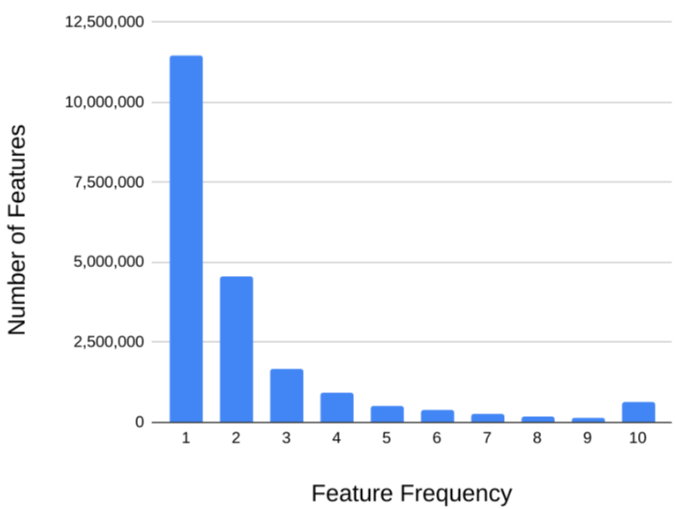}
   \caption{Features of SDK code are not uniformly distributed across a large set of SDKs. Features that appear in a single SDK are particularly useful for SDK matching and thus need to be weighed more in any similarity metric for SDK identification.}
   \label{fig:sdk-feature-freq}
\end{figure}

A preliminary data analysis showed that these features are non-uniformly distributed across SDKs. The leftmost bar in \autoref{fig:sdk-feature-freq} highlights how across a corpus of about 37,000 SDKs, there were about 11,000,000 features that occurred exactly in one SDK and thus could serve to make those SDKs uniquely reidentifiable in app code. As a result we add a weighing factor to each feature, to account for this non-uniformity:
\[
\textit{wfv}(m) = \textit{fv}(m)^T \times \textit{weights}.
\]

The high-dimensional space into which we represent Java methods naturally allows us to handle code composition. The vector representation of a group of methods (be they organized as a class, a package, an SDK, or other code structure) is the vector-sum of the corresponding vectors. We use \textit{cosine similarity} as our distance metric for two vectors $\textit{fv}_1$ and $\textit{fv}_2$:
\[
\mathit{\delta}(\textit{fv}_1, \textit{fv}_2) = \frac{ \textit{fv}_1 \textit{fv}_2 }{ \| \textit{fv}_1 \|\ \| \textit{fv}_2 \| }.
\]
Each class is assigned a corresponding vector representation as vector-sum of its component methods' vectors and by abuse of notation we write $\textit{fv}(c)$ for the vector of a class $c$, such that $\textit{fv}(c) = \Sigma_{m \in c} \textit{fv}(m)$. Then for each class in the app we search the set of all SDK classes for the classes that are minimally similar to the app class. Finally, we match SDKs that have sufficient support for being present in the app, measured by the number of classes in the SDK that are similar to app classes. \autoref{algo:mafia-scoring} presents the pseudo-code version of our technique.

The SDK-matching algorithm is parametrized by a threshold to ensure that app classes and SDK classes satisfy a minimum amount of similarity and by a second threshold to ensure an SDK is a match if a sufficient number of its classes appear in the app. In our experiments we use $0.2$ for the similarity threshold and $ 0.55$ for the class-count threshold. %

The representation of methods and classes as sparse vectors into a high-dimensional space allows us to borrow solutions to the Approximate Nearest Neighbor (ANN) problem, well studied in the information-retrieval domain. In particular a number of high-performance libraries such as Faiss~\cite{faiss}, SPTAG~\cite{sptag}, ScaNN~\cite{scann}, Hnswlib~\cite{hnswlib}, and NGT~\cite{ngt} provide capabilities to search for matches of thousands of classes in a typical app into a database of millions of SDK classes.

We evaluated the accuracy of this technique on a set of 302,397 SDKs for which we could find at least one mobile app that declared such an SDK as a dependency in its Gradle build configuration. The Gradle configuration data provided the ground truth against which we measured an average precision for SDK identification of $65.07\%$. Restricting to a smaller set of $9,683$ SDKs, the SDK identification reached $99.89\%$ precision at $46.16\%$ recall.  This implies that on this select set of SDKs, the identification technique can accurately detect the absence of an SDK from an app, though it may sometimes fail to detect the SDK's presence in an app.

In the context of the measurements in the rest of the paper, this application of the SDK-identification technique provides lower bounds on the market reach (cf. RQ3), since it undercounts the presence of SDKs of interest.

\begin{algorithm}[t]
    \DontPrintSemicolon
    
    \SetKwInOut{Input}{inputs}\SetKwInOut{Output}{outputs}
    \Input{\begin{tabular}[t]{ll}
            $A$ & \tcp{an app} \\
            $S_1, \dots, S_m$ & \tcp{a set of SDKs} \\
            $\eta$ & \tcp{class-dissimilarity upper bound} \\
            $\gamma$ & \tcp{SDK-similarity lower bound}
          \end{tabular}
          }
    \Output{\begin{tabular}[t]{l@{\qquad\qquad\ }l} $\mathcal{T}$ & \tcp{map from app classes to SDKs,} \\
                                                                     & \tcp{\quad $\mathcal{T} : A \mapsto 2^{S_1 \cup \dots \cup S_m}$} \end{tabular} }
    \BlankLine
    \Begin{
        let $A=\{c_1,\dots,c_n\}$ be the set of app classes \;
        let $\mathcal{C} = S_1 \cup \dots \cup S_m$ be the set of SDK classes \;
        \tcp{compute candidate SDKs for each app class }
        \ForEach{$c_i \in A$, $1 \leq i \leq n$}{
            $\mathcal{T}(c_i) \gets \left\{ c \in \mathcal{C} : \delta ( \mathit{fv}(c_i), \mathit{fv}(c^x_y) ) < \eta \right\}$ \;
        } \;
        \tcp{filter out SDKs with insufficient presence}
        \For{$j \gets 1$ \KwTo $m$}{
            \uIf{$ \left| S_j \cap \big( \mathcal{T}(c_1) \cup \dots \cup \mathcal{T}(c_n) \big) \right| \leq \gamma$}{
                \lFor{$i \gets 1$ \KwTo $n$}{$\mathcal{T}(c_i) \gets \mathcal{T}(c_i) \setminus S_j$}
            }
        }
    }
    \KwSty{end}\;
    
    \caption{SDK-Matching Algorithm}
    \label{algo:mafia-scoring}
\end{algorithm}

\section{Long-Form Definitions}
\label{sec:further_defs}

\begin{table*}[ht]

\begin{tabularx}{\textwidth}{@{} lX @{}}
\toprule
Category                   & Definition \\
\midrule
Ads                        & The SDK's main purpose is to support displaying ads, ads bidding, ads targeting, ads mediation, or user analytics with the express purpose of monetization or conversion. 

Examples: Mediation, ad adapters, ad push notifications  
\vspace{.5em} \\

\midrule
Analytics                  & 

\textit{Analytics (App Health)}

The SDK's main purpose is to track the systems performance of the application. 

Examples: crash logging, battery tracking and usage, UI latency

\vspace{.3em} 
\textit{Analytics (User Behavioral Analysis) }

The SDK's main purpose is to track user behavior on-device for business purposes (e.g. growth, retention, engagement, churn) without evidence of direct monetization or feedback into ads using the data collected. 

Ex. App engagement measurements 
\vspace{.5em} 
\\

\midrule
Security \& Authentication & 
The SDK's main purpose is to protect either the app or the user against malware and fraud.

\vspace{.3em} 
\textit{Security (Anti-Fraud)}

SDKs that provide fraud detection capabilities to an app, to handle account fraud, payment fraud, or identity fraud. The fraud detection may be performed on device or on a remote server.

\vspace{.3em} 
\textit{Security (Payments)}

SDKs that provide payment functionality, potentially managing all aspects of the payment flow (authorization, clearing, settlement, reversal). All payment methods are covered here.

\vspace{.3em} 
\textit{Security (Authentication)}

SDKs that authenticate the user using the app against a local or remote account or identity. All forms of authentication mechanisms (password, biometric, hardware token, knowledge based) are covered here.

\vspace{.3em} 
\textit{Security (Anti-malware)}

SDKs that attempt to scan for malware, bugs, or other on-device security issues.

\vspace{.3em} 
\textit{Security (Other) }

SDKs that do not clearly fit into one of the above.
\vspace{.5em} 
\\

\midrule
Tools~\slash~Other     & 

\textit{Location}

The SDK's main purpose is to  access location data and send it to a remote server.

\vspace{.3em} 
\textit{Location Tracking (Person)}

The SDK's main purpose is to track user location for the purpose of collecting business data (e.g. how often does a person visit a certain location, like Foursquare) or for the purpose of locating family members. 

\vspace{.3em} 
\textit{Location Tracking (Object)}

Specific object tracking (e.g. SDK that helps track a beacon / tag for finding lost objects) 

\vspace{.3em} 
\textit{Location Tracking (Maps) }

Any kind of navigational functionality, including the display of maps and map routes, and the location-based retrieval of points of interest.

\vspace{.3em} 
\textit{Location Tracking (Other)}

SDKs that do not clearly fit into one of the above 

\vspace{.3em} 
\textit{Social}

The SDK's main purpose is to connect the user to a larger social network, or provide a means of discovering other people, or provide a list of contacts for social engagement. Customer service chat SDKs are not included in the social category.

\vspace{.3em} 
\textit{Other}

The SDK's main purpose is to either support or provide functionality unrelated to any other defined categories. The SDK must have a purpose that is clearly defined in the information sources. 
\vspace{.5em} 
\\

\midrule
Unclear~\slash~Not found        & The SDK's main purpose is unclear using data from any of the allowed information sources, or, we cannot find any information on this SDK. 

Note that this category is distinct from ``Tools~\slash~Other'', which captures SDKs with clearly defined purposes.\\
\bottomrule

\end{tabularx}

\caption{Codebook definitions of categories created by our labeling process.}
\label{tab:codebook_full}

\end{table*}

Table \ref{tab:codebook_full} provides a full listing of the definitions developed and used in our labeling process. This table defines a classification system for Software Development Kits (SDKs) based on their primary functions, dividing them into five main categories: Ads, Analytics, Security and Authentication, Tools/Other, and Unclear/Not Found. The ``Ads'' category covers SDKs related to advertising; ``Analytics'' includes SDKs for tracking app health and user behavior; ``Security and Authentication'' encompasses SDKs for anti-fraud, payments, authentication, and anti-malware; ``Tools / Other'' includes location tracking (person, object, maps), social networking, and miscellaneous SDKs with defined but uncategorized purposes; and ``Unclear / Not Found'' is used when an SDK's purpose cannot be determined. Some categories are further broken down into subcategories with specific definitions and examples, such as ``Analytics (App Health)'' or ``Security (Anti-Fraud),'' to ensure precise classification.

Our coders used this information to categorize an SDK by first reviewing available information about the SDK's purpose and functionality. They then compared this information to the definitions provided in the table, starting with the main categories and then moving to the subcategories. By matching the SDK's features to the descriptions in the table, the rater assigned the most appropriate label. If the SDK's purpose is clearly defined but does not fit any existing category, they used the ``Other'' subcategory. If no information can be found, or the available information is insufficient to determine the SDK's purpose, the rater classified it as ``Unclear / Not Found.'' This systematic approach ensured consistent and accurate labeling of SDKs based on their primary function.

\section{List of Signal APIs}
\label{sec:API_LIST}

{
\newcolumntype{L}[1]{>{\raggedright\let\newline\\\arraybackslash\hspace{0pt}}p{#1}}

\small
\pgfplotstableset{every head row/.append style={before row={\caption{List of APIs used in fingerprinting-like behaviors observed in our SDK datasets.}
                                                               \label{tab:api-list} \\
                                                            \toprule
                                                            },
                                                after row={\midrule\endfirsthead
                                                           \caption[]{\textit{(cont'd)}} \\ \toprule Class Name & Property or Method Name \\ \midrule\endhead
                                                           \bottomrule\endfoot
                                                           },
                                                },
                 }
\pgfplotstabletypeset[row sep=\\, col sep=&, header=false, string type,
                      begin table=\onecolumn\begin{longtable},
                      end table=\end{longtable}\twocolumn,
                      columns/0/.style={column name={Class Name}, column type={p{0.48\textwidth }},
                                        preproc cell content/.append style={ /pgfplots/table/@cell content/.add={\lstinline!}{!}, },
                                        },
                      columns/1/.style={column name={Property or Method Name}, column type={L{0.48\textwidth }}, %
                                        string replace*={,}{!, \lstinline!},
                                        preproc cell content/.append style={ /pgfplots/table/@cell content/.add={\lstinline!}{!}, },
                                        },
                      outfile=pgfplotstable.example1.out.tex,
                      ]
{
android.accessibilityservice.AccessibilityServiceInfo & getResolveInfo(), getSettingsActivityName() \\
android.accounts.Account & name \\
android.accounts.AccountManager & getAccounts(), getAccountsByType(com.google) \\
android.app.ActivityManager & getDeviceConfigurationInfo(), getRunningAppProcesses(), getRunningTasks(), isUserAMonkey() \\
android.app.ActivityManager\$MemoryInfo & availMem, lowMemory, totalMem \\
android.app.ActivityManager\$RunningTaskInfo & numRunning \\
android.app.KeyguardManager & isDeviceSecure(), isKeyguardSecure() \\
android.app.TaskInfo & baseActivity(), numActivities(), taskId() \\
android.app.UiModeManager & getCurrentModeType() \\
android.app.WallpaperInfo & getPackageName() \\
android.app.WallpaperManager & getDrawable(), getWallpaperInfo() \\
android.app.admin.DevicePolicyManager & getActiveAdmins(), getStorageEncryptionStatus() \\
android.app.usage.StorageStatsManager & getTotalBytes() \\
android.app.usage.UsageStats & getPackageName() \\
android.app.usage.UsageStatsManager & queryUsageStats(INTERVAL\_DAILY) \\
android.bluetooth.BluetoothAdapter & getAddress(), getBondedDevices(), getDefaultAdapter(), getName(), getScanMode(), getState(), isDiscovering(), isEnabled() \\
android.content.ClipboardManager & getPrimaryClip(), getPrimaryClipDescription() \\
android.content.ComponentName & toShortString() \\
android.content.ContentResolver & query(Uri(content://com.google.android.gsf.gservices)), registerContentObserver(android.provider.MediaStore\$Images\$Media.EXTERNAL\_CONTENT\_URI) \\
android.content.Context & getPackageName() \\
android.content.Intent & getIntExtra(health), getIntExtra(plugged), getIntExtra(scale), getIntExtra(status), getIntExtra(temperature), getIntExtra(voltage), getStringExtra(technology) \\
android.content.pm.ApplicationInfo & flags, loadLabel(), sourceDir \\
android.content.pm.ConfigurationInfo & reqGlEsVersion \\
android.content.pm.InstallSourceInfo & getInstallingPackageName() \\
android.content.pm.PackageInfo & firstInstallTime, lastUpdateTime, packageName, receivers, requestedPermissions, requestedPermissionsFlags, services, signatures, signingInfo, versionCode, versionName \\
android.content.pm.PackageItemInfo & metaData, name, packageName \\
android.content.pm.PackageManager & checkPermission(), getApplicationLabel(), getInstallerPackageName(8), hasSystemFeature(android.hardware.fingerprint), hasSystemFeature(android.hardware.location.gps), hasSystemFeature(android.hardware.sensor.accelerometer), hasSystemFeature(android.hardware.sensor.compass), hasSystemFeature(android.hardware.sensor.light), hasSystemFeature(android.hardware.telephony), hasSystemFeature, hasSystemFeature, queryIntentActivities(), resolveActivity(android.content.Intent(``android.intent.action.CALL'')) \\
android.content.pm.SigningInfo & getApkContentsSigners() \\
android.content.res.Configuration & getLocales(), locale, orientation, screenLayout, uiMode \\
android.content.res.TypedArray & getDimensionPixelSize() \\
android.hardware.Sensor & getName(), getName(), getName(), getPower(), getVendor(), getVersion() \\
android.hardware.SensorEvent & values \\
android.hardware.SensorManager & getDefaultSensor(8) \\
android.hardware.camera2.CameraCharacteristics & CONTROL\_AE\_COMPENSATION\_RANGE, CONTROL\_AE\_LOCK\_AVAILABLE, CONTROL\_AF\_AVAILABLE\_MODES, CONTROL\_MAX\_REGIONS\_AF, FLASH\_INFO\_AVAILABLE, LENS\_FACING, REQUEST\_AVAILABLE\_CAPABILITIES, SCALER\_AVAILABLE\_MAX\_DIGITAL\_ZOOM, SCALER\_STREAM\_CONFIGURATION\_MAP,  SENSOR\_INFO\_SENSITIVITY\_RANGE, STATISTICS\_INFO\_MAX\_FACE\_COUNT \\
android.hardware.camera2.CameraManager & getCameraIdList() \\
android.hardware.camera2.params.StreamConfigurationMap & getHighSpeedVideoFpsRanges(), getOutputSizes() \\
android.hardware.usb.UsbManager & getDeviceList() \\
android.location.Location & getAccuracy(), getAltitude(), getBearing(), getBearingAccuracyDegrees(), getElapsedRealtimeNanos(), getLatitude(), getLongitude(), getProvider(), getSpeed(), getSpeedAccuracyMetersPerSecond(), getTime(), getVerticleAccuracyMeters(), isFromMockProvider() \\
android.location.LocationManager & getBestProvider(), getLastKnownLocation(), isProviderEnabled(gps), isProviderEnabled(network) \\
android.media.AudioManager & getDevices(), getRingerMode(), getStreamMaxVolume(), getStreamVolume(), isMusicActive() \\
android.media.MediaDrm & getPropertyByteArray(deviceUniqueId) \\
android.media.RingtoneManager & getActualDefaultRingtoneUri(), getDefaultUri(), getRingtone() \\
android.net.ConnectivityManager & getActiveNetwork(), getActiveNetworkInfo(), getDefaultProxy(), getNetworkCapabilities() \\
android.net.NetworkCapabilities & hasTransport(4) \\
android.net.NetworkInfo & getState(), getTypeName(), isConnected(), isRoaming() \\
android.net.ProxyInfo & getHost() \\
android.net.TrafficStats & getTotalRxBytes(), getTotalTxBytes() \\
android.net.sip.SipManager & isVoipSupported() \\
android.net.wifi.ScanResult & capabilities(), frequency(), level(), SSID() \\
android.net.wifi.WifiInfo & getBSSID(), getFrequency(), getIpAddress(), getLinkSpeed(), getMacAddress(), getNetworkId(), getRssi(), getSSID() \\
android.net.wifi.WifiManager & calculateSignalLevel(), getConfiguredNetworks(), getConnectionInfo(), getScanResults(), getSSID(), is5GHzBandSupported(), isDeviceToApRttSupported(), isEnhancedPowerReportingSupported(), isP2pSupported(), isPreferredNetworkOffloadSupported(), isScanAlwaysAvailable(), isTdlsSupported(), isWifiEnabled() \\
android.os.BaseBundle & getBoolean(present), getLong(install\_begin\_timestamp\_seconds), getLong(referrer\_click\_timestamp\_seconds), getString(technology), getString(install\_referrer) \\
android.os.Build & BOARD, BOOTLOADER, BRAND, CPU\_ABI, CPU\_ABI2, DEVICE, DISPLAY, FINGERPRINT, getRadioVersion(), getSerial(), HARDWARE, HOST, ID, MANUFACTURER, MODEL, PRODUCT, RADIO, SERIAL, SOC\_MANUFACTURER, SOC\_MODEL, SUPPORTED\_32\_BIT\_ABIS, SUPPORTED\_64\_BIT\_ABIS, SUPPORTED\_ABIS, TAGS, TIME, TYPE, USER \\
android.os.Build\$VERSION & BASE\_OS, CODENAME, INCREMENTAL, RELEASE, SDK\_INT, SECURITY\_PATCH \\
android.os.Build\$VERSION\_CODE & FROYO, GINGERBREAD\_MR1, GINGERBREAD, HONEYCOMB\_MR1, HONEYCOMB\_MR2, HONEYCOMB, ICE\_CREAM\_SANDWICH\_MR1, ICE\_CREAM\_SANDWICH, JELLY\_BEAN\_MR1, JELLY\_BEAN\_MR2, JELLY\_BEAN, KITKAT\_WATCH, KITKAT, LOLLIPOP\_MR1, LOLLIPOP \\
android.os.Debug & isDebuggerConnected() \\
android.os.Environment & getDataDirectory(), getExternalStorageDirectory(), getExternalStorageState(), getRootDirectory(), isExternalStorageEmulated() \\
android.os.PowerManager & getCurrentThermalStatus(), getLocationPowerSaveMode(), isDeviceIdleMode(), isInteractive(), isPowerSaveMode() \\
android.os.StatFs & getAvailableBlocks(), getAvailableBlocksLong(), getBlockCount(), getBlockCountLong(), getBlockSize(), getBlockSizeLong(), getTotalBytes() \\
android.os.SystemClock & elapsedRealtime(), uptimeMillis() \\
android.os.SystemProperties & get(gsm.operator.isroaming), get(gsm.operator.numeric), get(gsm.sim.state), get(init.svc.qemu\-props), get(init.svc.qemud), get(qemu.hw.mainkeys), get(qemu.sf.fake\_camera), get(qemu.sf.lcd\_density), get(ro.bootloader), get(ro.bootmode), get(ro.hardware), get(ro.kernel.android.qemud), get(ro.kernel.qemu.gles), get(ro.product.device), get(ro.product.model), get(ro.product.name), get(ro.runtime.firstboot), get(ro.serialno) \\
android.os.UserManager & getSerialNumberForUser(), getUserProfiles(), isDemoUser(), isSystemUser(), supportsMultipleUsers() \\
android.os.storage.StorageVolume & isPrimary() \\
android.provider.Settings\$Global & getString(adb\_enabled), getString(airplane\_mode\_on), getString(airplane\_mode\_radios), getString(always\_finish\_activities), getString(animator\_duration\_scale), getString(auto\_time), getString(auto\_time\_zone), getString(bluetooth\_discoverability), getString(bluetooth\_discoverability\_timeout), getString(bluetooth\_on), getString(boot\_count), getString(data\_roaming), getString(development\_settings\_enabled), getString(device\_provisioned), getString(http\_proxy), getString(mode\_ringer), getString(network\_preference), getString(stay\_on\_while\_plugged\_in), getString(transition\_animation\_scale), getString(usb\_mass\_storage\_enabled), getString(use\_google\_mail), getString(wait\_for\_debugger), getString(wifi\_networks\_available\_notification\_on) \\
android.provider.Settings\$Secure & getInt(accessibility\_enabled), getInt(mock\_location), getString(accessibility\_display\_inversion\_enabled), getString(allowed\_geolocation\_origins), getString(default\_input\_method), getString(enabled\_input\_methods), getString(input\_method\_selector\_visibility), getString(install\_non\_market\_apps), getString(location\_mode), getString(skip\_first\_use\_hints), getString(tts\_default\_pitch), getString(tts\_default\_rate), getString(tts\_default\_synth), getString(tts\_enabled\_plugins) \\
android.provider.Settings\$System & getInt(screen\_brightness), getString(accelerometer\_rotation), getString(android\_id), getString(auto\_caps), getString(auto\_punctuate), getString(auto\_replace), getString(dtmf\_tone), getString(dtmf\_tone\_type), getString(end\_button\_behavior), getString(font\_scale), getString(haptic\_feedback\_enabled), getString(mode\_ringer\_streams\_affected), getString(mute\_streams\_affected), getString(notification\_sound), getString(ringtone), getString(screen\_brightness), getString(screen\_brightness\_mode), getString(screen\_off\_timeout), getString(show\_password), getString(sound\_effects\_enabled), getString(time\_12\_24), getString(user\_rotation), getString(vibrate\_on), getString(vibrate\_when\_ringing) \\
android.provider.Telephony\$Sms & getDefaultSmsPackage() \\
android.system.StructStat & st\_mtime \\
android.telecom.TelecomManager & isTtySupported() \\
android.telephony.CellIdentityGsm & getCid(), getLac(), getMcc(), getMnc() \\
android.telephony.CellIdentityLte & getCi(), getMcc(), getMnc(), getTac() \\
android.telephony.CellIdentityWcdma & getCid(), getLac(), getMcc(), getMnc() \\
android.telephony.CellInfo & isRegistered() \\
android.telephony.CellInfoCdma & getCellIdentity(), getCellSignalStrength() \\
android.telephony.CellInfoGsm & getCellIdentity(), getCellSignalStrength() \\
android.telephony.CellInfoLte & getCellIdentity(), getCellSignalStrength() \\
android.telephony.CellInfoTdscdma & getCellSignalStrength() \\
android.telephony.CellInfoWcdma & getCellIdentity(), getCellSignalStrength() \\
android.telephony.CellSignalStrength & getDbm() \\
android.telephony.CellSignalStrengthCdma & getDbm() \\
android.telephony.CellSignalStrengthGsm & getDbm() \\
android.telephony.CellSignalStrengthLte & getDbm() \\
android.telephony.CellSignalStrengthTdscdma & getDbm() \\
android.telephony.CellSignalStrengthWcdma & getDbm() \\
android.telephony.SignalStrength & getCellSignalStrengths() \\
android.telephony.SubscriptionInfo & getCarrierName(), getCountryIso(), getDataRoaming(), getDisplayName(), getIccId(), getNumber(), getSimSlotIndex() \\
android.telephony.TelephonyManager & getActiveModemCount(), getCellLocation(), getDataNetworkType(), getDataState(), getDeviceId(), getDeviceSoftwareVersion(), getGroupIdLevel1(), getImei(), getLine1Number(), getMeid(), getMmsUAProfUrl(), getMmsUserAgent(), getNetworkCountryIso(), getNetworkOperator(), getNetworkOperatorName(), getNetworkType(), getPhoneCount(), getPhoneType(), getSignalStrength(), getSimCountryIso(), getSimOperator(), getSimOperatorName(), getSimSerialNumber(), getSimSpecificCarrierIdName(), getSimState(), getSubscriberId(), getVoiceMailAlphaTag(), getVoiceMailNumber(), hasIccCard(), isHearingAidCompatibilitySupported(), isNetworkRoaming(), isSmsCapable(), isVoiceCapable(), isWorldPhone() \\
android.telephony.cdma.CdmaCellLocation & getBaseStationId(), getNetworkId(), getSystemId() \\
android.telephony.gsm.GsmCellLocation & getCid(), getLac() \\
android.util.DisplayMetrics & density, densityDpi, heightPixels, scaledDensity, widthPixels, xdpi, ydpi \\
android.view.Display & getMetrics(), getName(), getRealMetrics(), getRefreshRate(), getRotation(), getSize() \\
android.view.accessibility.AccessibilityManager & getEnabledAccessibilityServiceList(), getInstalledAccessibilityServiceList(), isTouchExplorationEnabled() \\
android.view.inputmethod.InputMethodInfo & getPackageName() \\
android.webkit.WebSettings & getDefaultUserAgent(), getUserAgentString() \\
android.widget.TextView & getTextSize() \\
androidx.ads.identifier.AdvertisingIdClient & getAdvertisingIdInfo(), isAdvertisingIdProviderAvailable() \\
androidx.ads.identifier.AdvertisingIdInfo & getId() \\
androidx.core.hardware.fingerprint.FingerprintManagerCompat & isHardwareDetected() \\
androidx.core.location.LocationManagerCompat & isLocationEnabled() \\
com.google.android.gms.ads.identifier.AdvertisingIdClient\$Info & getId() \\
com.google.android.gms.common.GoogleApiAvailability & isGooglePlayServicesAvailable() \\
com.google.android.gms.location.FusedLocationProviderClient & getLastLocation() \\
com.google.android.gms.location.LocationResult & getLastLocation() \\
com.scottyab.rootbeer.RootBeer & isRooted(), isRootedWithoutBusyBoxCheck() \\
java.io.File & (/system/fonts/), getFreeSpace(), getTotalSpace() \\
java.io.FileReader & (/proc/version) \\
java.io.RandomAccessFile & (/proc/meminfo), (/sys/devices/system/cpu/cpu0/cpufreq/stats/time\_in\_state), (sys/devices/system/cpu/cpu0/cpufreq/cpuinfo\_max\_freq), (sys/devices/system/cpu/cpu0/cpufreq/cpuinfo\_min\_freq) \\
java.lang.Class & forName(android.os.SystemProperties), forName(com.google.android.gms.loca\-tion.FusedLocationProviderClient), forName(dalvik.system.Taint), forName(androidx.ads.identifier.AdvertisingIdClient), forName(com.google.android.gms.ads.iden\-tifier.AdvertisingIdClient), getDeclaredField(sSystemFontMap), getDeclaredFields(), getDeclaredMethod(getFatVolumeId), getDeclaredMethods(), getMethod(getEnrolledFingerprints), getMethod(getFingerId) \\
java.lang.Runtime & availableProcessors(), maxMemory() \\
java.lang.System & currentTimeMillis(), getProperty(http.agent), getProperty(http.proxyHost), getProperty(http.proxyPort), getProperty(os.arch), getProperty(os.name), getProperty(os.version) \\
java.lang.Thread & getStackTrace() \\
java.net.InetAddress & getHostAddress() \\
java.net.NetworkInterface & getHardwareAddress(), getHostAddress(), getMTU(), getName(), isUp() \\
java.security.KeyStore & getCertificate(), getCreationDate() \\
java.security.cert.X509Certificate & getIssuerDN(), getKeyUsage() \\
java.util.Calendar & getCalendarType(), getTime(), getTimeInMillis() \\
java.util.Locale & getCountry(), getDefault(), getDisplayCountry(), getDisplayLanguage(), getDisplayName(), getDisplayVariant(), getLanguage(), getScript(), toString(), US() \\
java.util.TimeZone & getDefault(), getDisplayName(), getDSTSavings(), getId(), getOffset(), getRawOffset(), inDaylightTime(), useDaylightTime() \\
}
}

\end{document}